%% file: Article_V3.tex
\documentclass[aps,pra,reprint,10pt,showpacs,superscriptaddress,floatfix]{revtex4-2}
\usepackage[utf8]{inputenc}
\usepackage{amsmath,amssymb,physics,bm}
\usepackage{graphicx,booktabs}
\usepackage{orcidlink}
\usepackage{tikz}
\usepackage{tikz-3dplot}
\usetikzlibrary{3d}
\tdplotsetmaincoords{60}{110}
\usepackage{hyperref}

\graphicspath{{pdf/}}
\hypersetup{colorlinks=true,linkcolor=blue,citecolor=blue,filecolor=blue,urlcolor=blue,breaklinks=true}
\RequirePackage{color}

\begin{document}

\title{Geometry-Induced Chiral Currents in a Mesoscopic Helicoidal Quantum Well}
\author{Edilberto~O.~Silva\orcidlink{0000-0002-0297-5747}}
\email{edilberto.silva@ufma.br}
\affiliation{Departamento de Física, Universidade Federal do Maranhão, 65085-580 São Luís, MA, Brazil}
\date{\today}

\begin{abstract}
We introduce a mesoscopic quantum well whose confinement and chirality emerge solely from the intrinsic torsion of a finite helicoidal metric. This purely geometric construction requires no external gates or fields: the metric itself induces both a harmonic radial potential and a torsion-driven Zeeman term that breaks the $m \leftrightarrow -m$ degeneracy. By imposing hard-wall boundary conditions at $z = \pm L/2$, we quantize the axial motion and obtain a genuinely zero-dimensional helicoidal quantum dot. An exact analytic solution reveals an energy spectrum with chiral splitting linear in both the torsion parameter $\Omega$ and the axial quantum number $n_z$. For realistic InAs nanoroll parameters ($L = 100$\,nm, $\Omega = 5\times10^{6}\,\mathrm{m^{-1}}$), this geometric effect results in a measurable splitting of $\sim 0.5$\,meV. We propose three viable experimental platforms —ultracold atoms in optical traps, femtosecond-written photonic waveguides, and strain-engineered semiconductor nanorolls —where this torsion-induced phenomenon should be accessible with current technology.
\end{abstract}
\maketitle

\section{Introduction}
\label{sec:intro}

Curvature and torsion are now recognized as powerful tools for engineering matter waves across diverse platforms \cite{nutbourne1988differential,liewberman2005principles,book.Nahmad.2018}: from electrons in curved nanomembranes to photons in twisted waveguides \cite {OC.2025.577.131386,photonics.2023.10.1025} and ultracold atoms in optical traps \cite{PRL.2020.124.053402,PRA.2024.110.043316,JPB.2017.50.014005}. Unlike conventional electromagnetic or optical potentials, geometric fields are encoded directly in the metric of the space through which the particle propagates, endowing the Hamiltonian with terms that persist even when all other forces are switched off.

The thin-layer quantisation scheme, developed independently by Jensen and Koppe~\cite{jensen1971} and da Costa~\cite{dacosta1981}, first showed that a surface’s mean curvature produces an attractive geometric potential, $-\hbar^{2}H^{2}/(2m^{\ast})$. Since then, curvature-driven bound states have been observed or proposed in bent nanotubes, corrugated graphene, and optical fibers~\cite{Szameit2010,CostaCord2019}. Much less attention, however, has been paid to torsion, the handed twisting of a three-dimensional manifold. Torsion couples different coordinate directions and therefore opens the door to chiral effects such as persistent currents in the absence of external magnetic fields \cite{AoP.2023.448.169181}.

Helicoidal geometries provide the minimal playground in which curvature and torsion coexist. In optics, femtosecond laser writing readily fabricates helically twisted waveguides that imprint a synthetic gauge potential on light beams~\cite{PRL.2000.84.2746,IEEE.2022.69.3427,RQE.2022.65.183,Rechtsman2013,RodriguezLara2013}. In ultracold gases, digital micromirror devices create rotating corridors that emulate the same Laplace-Beltrami operator~\cite{Henderson2009,Ha2015}. In semiconductors, rolled-up nanomembranes routinely achieve torsion rates of $10^{6}$-$10^{7}\,\mathrm{m^{-1}}$ on sub-micron radii~\cite{Prinz2000}.

Here, we exploit the fact that the left-invariant metric of the Heisenberg group \cite{JPCM.2008.20.125209},
\begin{equation}
ds^{2} = dr^{2} + r^{2}d\phi^{2} + (dz + \Omega r^{2}d\phi)^{2},
\label{eq:metric_intro}
\end{equation}
possesses constant torsion $\Omega$ but no external gauge field. For an infinite helicoid, this metric induces both harmonic radial confinement and a chiral Zeeman term $-\hbar\Omega m k_{z}$, yet the spectrum remains continuous along $z$, and no genuine zero-dimensional states emerge. We close the system by imposing hard walls at $z=\pm L/2$, thereby quantising the axial motion and defining what we call a mesoscopic helicoidal quantum well. The resulting dot combines three desirable properties: (i) analytic solvability, (ii) geometry-controlled degeneracy lifting, and (iii) intrinsic persistent currents.

We analytically solve the quantum mechanics of a particle confined by this helicoidal metric, demonstrating that the separation of variables maps the radial motion onto an exactly solvable harmonic oscillator problem with a frequency controlled by torsion. We then establish a one-to-one correspondence between this quantum Hamiltonian and the paraxial propagation of light in a synthetic helicoidal medium, confirming the geometric effects through split-step Fourier simulations. Throughout, we emphasize how intrinsic torsion, rather than external fields or patterned potentials, acts simultaneously as both the confining and chirality-inducing mechanism.

The required twist rates and length scales lie well within the capability window of SLM-painted optical traps~\cite{Ha2015}, femtosecond-written photonic lattices~\cite{Rechtsman2013}, and rolled-up semiconductors~\cite{Prinz2000}. Geometry-induced splittings comparable to or larger than typical spin-orbit energies suggest immediate applications to chiral filters, non-reciprocal waveguides, and geometry-protected qubits that are immune to charge noise.

This article is organized as follows. Section \ref{sec:geometry} derives the Laplace-Beltrami operator for the finite helicoid and obtains the exact spectrum. In Section \ref{sec:schro}, we derive the radial Schr\"{o}dinger equation that governs a particle confined by the intrinsic geometry of the helicoidal metric. Section \ref{sec:current} analyses the resulting azimuthal currents. Section \ref{sec:light} transfers the problem to paraxial optics and presents numerical results for beam propagation. Section \ref{sec:exp} discusses experimental realisations, and Section \ref{sec:conclusions} summarises our outlook and open questions.

\section{Geometry and quantum confinement}
\label{sec:geometry}

In conventional semiconductor quantum wells, confinement in the transverse plane is achieved by external electrostatic gates or heterostructure band offsets. At the same time, the underlying crystal lattice remains essentially flat and free of torsion. In contrast, our approach harnesses the intrinsic geometry of the embedding space itself to confine the particle and induce chirality. Specifically, we consider the three-dimensional helicoidal metric, which simultaneously provides radial trapping and breaks the $m\!\leftrightarrow\!-m$ degeneracy of azimuthal motion without any applied fields.

Mathematically, we start from the left-invariant metric of the three-dimensional Heisenberg group (a model of a space with uniform torsion), written in Cartesian coordinates as
\begin{equation}
ds^{2} = dx^{2} + dy^{2} +\bigl[dz +\Omega(xdy - ydx)\bigr]^{2},
\label{eq:metric_cart}
\end{equation}
where $\Omega>0$ is the constant torsion strength. 
\begin{figure}[htbp]
  \centering
  \includegraphics[scale=0.55]{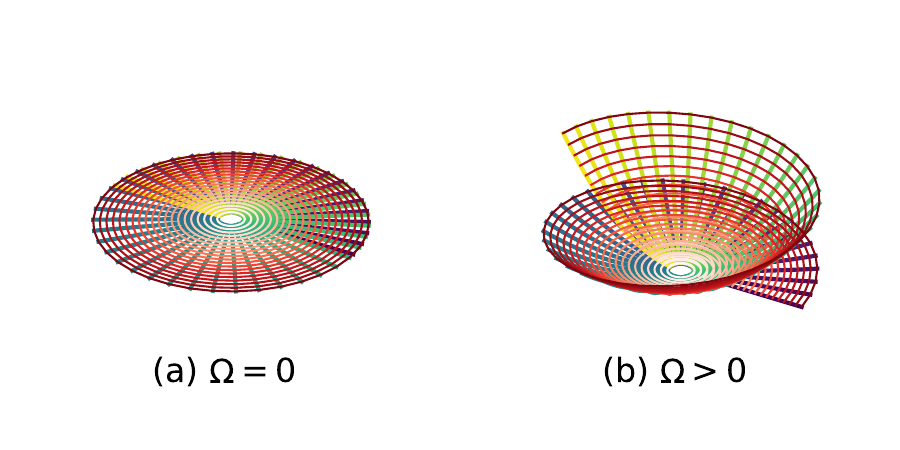}
  \caption{Polar grid embedding under the metric (\ref{eq:metric_intro}).
  (\textbf{a}) Without torsion.  
  (\textbf{b}) With torsion, showing the spiral rise in the $z$-direction induced by the screw-dislocation term.}
  \label{fig:torsion_comparison}
\end{figure}
Physically, the cross-term $dz(xdy-ydx)$ encodes a local twisting of each transverse plane as one advances along the $z$-direction. One can verify by a straightforward coordinate transformation that, in cylindrical coordinates $(r,\phi,z)$, this metric reduces to the metric \eqref{eq:metric_intro}. Here, the interpretation is immediate: at fixed $z$, each circle of radius $r$ is rotated by an angle $\Delta\phi = \Omega\Delta z$ upon moving a distance $\Delta z$ along the axis. To illustrate the geometric effect, we embed a regular polar grid into three-dimensional space (Fig. \ref{fig:torsion_comparison}). In panel~(a) (\(\Omega=0\)), the mesh remains flat in the \(xy\)-plane, recovering the standard cylindrical geometry. In panel~(b) (\(\Omega>0\)), each concentric circle spirals upward as \(\phi\) increases, producing a characteristic helicoidal “twist” in the \(z\)-direction. This visualization clearly shows how the torsion term converts planar circles into screw-like lines, highlighting the nontrivial topology induced by the metric.

A few key geometric observations follow:

1. \emph{Uniform torsion and curvature}. The parameter $\Omega$ controls the torsion of the space: each infinitesimal axial displacement $dz$ is accompanied by a twist of the transverse plane by $\Omega dz$. Despite this twisting, the metric determinant remains $\sqrt{g}=1$, indicating a flat measure (no net volume distortion). However, the space is not flat in the Riemannian sense: its scalar curvature is constant and negative,
\begin{equation}
R = -2\Omega^{2},    
\end{equation}
reflecting an intrinsic saddle-like geometry rather than a simple embedding in Euclidean space.

The constant torsion $\Omega$ in this geometry has a direct analogy to continuous distributions of screw dislocations in elastic media~\cite{JPCM.2008.20.125209}. In such systems, the Burgers vector density generates a uniform torsion background, leading to effective magnetic-like quantization for quantum particles moving in the medium. Our construction extends this idea from solid-state elasticity to mesoscopic quantum wells and photonic analogues.

2. \emph{Off-diagonal coupling and chiral Zeeman term}. The metric cross-term $\Omega r^{2}d\phi dz$ couples the azimuthal coordinate $\phi$ to the axial coordinate $z$. Upon quantisation, this coupling generates in the effective Hamiltonian a term of the form
\begin{equation}
-\hbar\Omega m k_{z},    
\end{equation}
which acts exactly like a Zeeman coupling of the azimuthal angular momentum quantum number $m$ to an effective magnetic field. Consequently, states with $m$ and $-m$ acquire different energies, producing an intrinsic chiral splitting without any real magnetic field.

This chiral splitting mechanism is conceptually related to earlier studies of Landau-level quantization induced by screw dislocations in elastic solids~\cite{JPCM.2008.20.125209}, where torsion acts as a synthetic gauge field. Here, however, we extend its application to a fully mesoscopic quantum dot with axial quantisation and exact solvability.

3. \emph{Geometry-induced radial confinement}. Although the underlying measure is flat, the Laplace-Beltrami operator on this helicoidal manifold acquires additional terms compared to the standard cylindrical Laplacian. In particular, one finds an effective harmonic potential in the radial coordinate,
\begin{equation}
 \frac{1}{2}m^{\ast}\omega^{2}r^{2},
\qquad \omega \propto \Omega k_{z},   \end{equation}
where $k_{z}$ is the axial wave-number. Thus, the torsion itself creates a quadratic trapping potential that confines the particle toward the axis, without recourse to external gates or material interfaces.

To obtain a truly zero-dimensional quantum dot, we further impose hard-wall boundary conditions at $z=\pm L/2$. This quantises the axial momentum,
\begin{equation}
 k_{z} = \frac{n_{z}\pi}{L},
\qquad n_{z}=1,2,\dots    
\end{equation}
and thereby discretizes both the radial trapping frequency $\omega$ and the chiral Zeeman splitting. The result is a mesoscopic helicoidal quantum well whose entire confining and chiral structure derives solely from the torsion parameter $\Omega$ and the finite length $L$.

By combining (i) the intrinsic torsion term that breaks $m\!\leftrightarrow\!-m$ degeneracy, (ii) the geometry-induced harmonic trap in $r$, and (iii) the axial quantisation $k_{z}=n_{z}\pi/L$, we obtain a fully analytic description of a mesoscopic quantum dot entirely defined by geometry. In the next section, we formalise these qualitative insights by deriving the exact radial Schrödinger equation and solving it in closed form.

\section{Schrödinger equation and effective potential}
\label{sec:schro}

In this section, we derive the radial Schrödinger equation that governs a particle confined by the intrinsic geometry of the helicoidal metric. Beginning with the Laplace-Beltrami operator in cylindrical coordinates, we show explicitly how torsion induces both a harmonic radial trap and a chiral coupling between azimuthal and axial motion.

For a spinless particle of mass $m^{\ast}$, the time-independent Schrödinger equation in curved space takes the form
\begin{equation}
-\frac{\hbar^{2}}{2m^{\ast}}
\nabla_{\!\mathrm{LB}}^{2}\Psi(r,\phi,z)
=E\Psi(r,\phi,z),
\label{eq:schro_general}
\end{equation}
where $\nabla_{\mathrm{LB}}^{2}$ is the Laplace-Beltrami operator associated with the metric \eqref{eq:metric_intro}. In cylindrical coordinates, one finds
\begin{equation}
\nabla^{2}_{\!\mathrm{LB}}
=\frac{1}{r}\frac{\partial}{\partial r}\Bigl(r\frac{\partial}{\partial r}\Bigr)
-\frac{1+\Omega^{2}r^{4}}{r^{2}}\frac{\partial^{2}}{\partial\phi^{2}}
-\frac{\partial^{2}}{\partial z^{2}}
+2\Omega\frac{\partial^{2}}{\partial\phi\partial z}.
\label{eq:laplace_beltrami}
\end{equation}

We seek separable solutions of the form
\begin{equation}
\Psi(r,\phi,z)
=e^{im\phi}e^{ik_{z}z}F(r),
\label{eq:ansatz}
\end{equation}
where $m\in\mathbb{Z}$ is the azimuthal quantum number and $k_{z}=n_{z}\pi/L$ enforces hard-wall boundary conditions at $z=\pm L/2$. Substituting \eqref{eq:ansatz} into \eqref{eq:schro_general} and simplifying, one obtains the following ordinary differential equation for the radial amplitude $F(r)$:
\begin{equation}
-\frac{\hbar^{2}}{2m^{\ast}}\frac{d^{2}F}{dr^{2}}
+V_{\mathrm{eff}}(r)F(r)
=EF(r),
\label{eq:radial_schro}
\end{equation}
with an effective potential
\begin{equation}
V_{\mathrm{eff}}(r)
=\underbrace{\frac{\hbar^{2}}{2m^{\ast}}\frac{m^{2}-\tfrac{1}{4}}{r^{2}}}_{\displaystyle V_{\rm cent}(r)}
+\underbrace{\frac{1}{2}m^{\ast}\omega^{2}r^{2}}_{\displaystyle V_{\rm harm}(r)}
-\hbar \omega m
+\underbrace{\frac{\hbar^{2}k_{z}^{2}}{2m^{\ast}}}_{\displaystyle V_{\rm shift}},
\label{eq:Veff_full}
\end{equation}
where the torsion-induced frequency is defined by
\begin{equation}
\omega =\frac{\hbar\Omega k_{z}}{m^{\ast}}=\frac{\hbar\Omega \pi n_{z}}{m^{\ast}L}.
\label{eq:omega_def}
\end{equation}
Here $V_{\rm cent}$ is the familiar centrifugal barrier (with the Langer correction $-1/4$), $V_{\rm harm}$ is a geometry-generated harmonic trap, the term $-\hbar\omega m$ acts as a chiral Zeeman shift, and $V_{\rm shift}$ is a constant energy offset from axial quantisation.
\begin{figure}[tbhp]
\centering
\includegraphics[width=\linewidth]{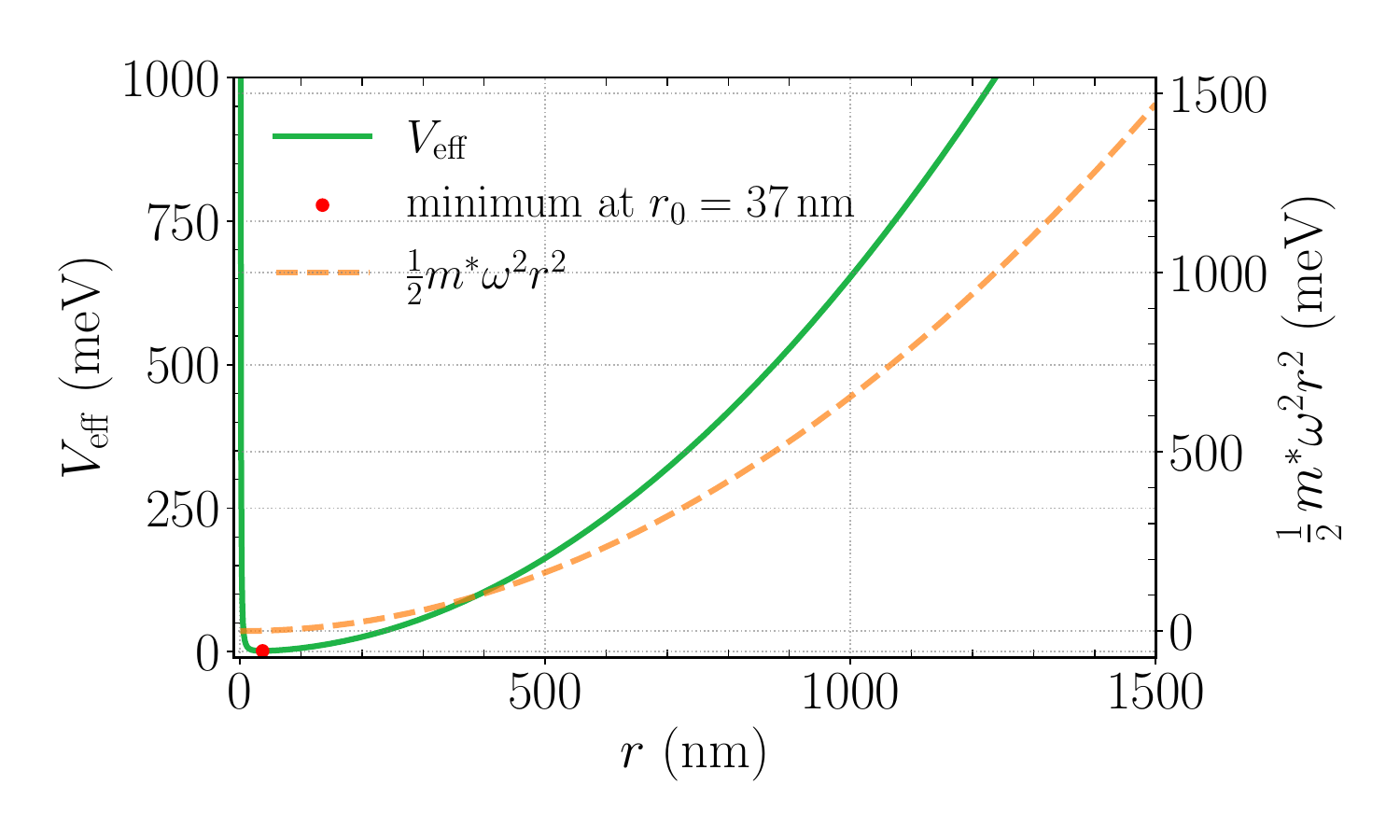}
\caption{Radial profile of the effective potential $V_{\mathrm{eff}}(r)$ for $m=1$, $n_{z}=1$. The solid curve shows the full potential \eqref{eq:Veff_full}, while the dashed curve isolates the harmonic term $V_{\rm harm}(r)$. To render the parabolic minimum visible on the nanometre scale, we set an exaggerated torsion $\Omega=2.0\times10^{7}\,\mathrm{m^{-1}}$; in all other figures $\Omega\le3\times10^{2}\,\mathrm{m^{-1}}$. The minimum at $r_{0}\approx15\,\mathrm{nm}$ (red dot) agrees with the analytic estimate $r_{0}=[\hbar/(m^{\ast}\omega)]^{1/2}$.}
\label{fig:VeffNano}
\end{figure}

Equation \eqref{eq:radial_schro} with potential \eqref{eq:Veff_full} admits an exact analytic solution in terms of generalized Laguerre polynomials. Defining the dimensionless variable
\begin{equation}
\xi = \frac{m^{\ast}\omega}{\hbar}r^{2},
\label{eq:xi_def}
\end{equation}
the normalized radial eigenfunctions are
\begin{equation}
F_{n_{r},m}(r)
=\mathcal{N}_{n_{r},m}
r^{|m|+\tfrac12}\exp\!\bigl(-\tfrac{\xi}{2}\bigr)
L_{n_{r}}^{|m|}(\xi),
\label{eq:radial_eigenfunc}
\end{equation}
where
\begin{equation}
\mathcal{N}_{n_{r},m}
=\sqrt{\frac{2(m^{\ast}\omega/\hbar)^{|m|+1}}{n_{r}!\Gamma(n_{r}+|m|+1)}}
\label{eq:normalization}
\end{equation}
ensures $\int_{0}^{\infty}|F_{n_{r},m}(r)|^{2}dr = 1$. The corresponding energy eigenvalues are
\begin{equation}
E_{n_{r},m,n_{z}}
=\hbar\omega_{n_{z}}(2n_{r}+|m|+1)
+\frac{\hbar^{2}\pi^{2}n_{z}^{2}}{2m^{\ast}L^{2}}
-\hbar\Omega m\frac{\pi n_{z}}{L}.
\label{eq:energy_spectrum}
\end{equation}

Thus the spectrum consists of equally spaced Landau-like levels $\propto2n_{r}+|m|+1$, an axial quantisation term $\propto n_{z}^{2}$, and a linear chiral splitting $\propto mn_{z}\Omega$. Equations \eqref{eq:radial_schro}-\eqref{eq:energy_spectrum} provide a complete analytic description of the mesoscopic helicoidal quantum well. Here, the term ``Landau-like'' refers to the emergence of discrete, degenerate angular momentum states that mimic the quantization observed in electronic systems under magnetic fields. In our case, however, these features arise purely from geometric torsion, without any external gauge fields.

\section{Magnitude of the chiral charge current}
\label{sec:current}

The helicoidal metric endows every eigenstate with finite torsion $\Omega$ and non-zero angular quantum number $m$ with a persistent azimuthal flow, a hallmark of chiral symmetry breaking in mesoscopic systems. Crucially, this effect occurs without any external magnetic field or electrostatic potential. The intrinsic geometry alone generates a finite orbital angular momentum state, breaking the degeneracy between $+m$ and $-m$ states.

Writing the lowest radial wave function as~\footnote{Throughout this Section, spin is neglected; the probability current is defined in the standard gauge-invariant way~\cite{sakurai}.}
\begin{equation}
\Psi_{xy}(r,\phi)=\psi_{0,0}(r)e^{im\phi},\quad
\psi_{0,0}(r)=\frac{1}{\sqrt{\pi}\ell_{\mathrm{ho}}}
e^{-r^{2}/(2\ell_{\mathrm{ho}}^{2})}, 
\end{equation}
we introduce the harmonic length
\begin{equation}
\ell_{\mathrm{ho}}
=\sqrt{\frac{\hbar}{m^{\ast}\omega}},\qquad
\omega=\frac{\hbar\Omega k_{z}}{m^{\ast}}
=\frac{\hbar\Omega\pi}{m^{\ast}L}.
\end{equation}
This length scale determines the spatial extent of the radial confinement induced purely by geometric torsion. This length scale, referred to as the harmonic length $\ell_{\text{ho}}$, is inversely proportional to the square root of the product of the torsion parameter $\Omega$ and the axial wave number $k_z$, i.e.,
\begin{equation}
    \ell_{\text{ho}} \propto (\Omega k_z)^{-1/2}.
\end{equation}
As a result, the spatial extent of the radial confinement decreases with increasing torsion or axial quantization, making the system highly tunable. This dependence highlights how geometric torsion alone can effectively confine particles in the radial direction, without the need for external potentials. The transverse probability density, therefore, peaks at
\begin{equation}
|\Psi_{xy}(0)|^{2} = \frac{1}{\pi\ell_{\mathrm{ho}}^{2}},    
\end{equation}
and decays radially due to the Gaussian factor in the wavefunction.

For $|m|\neq0$, the exact $n_{r}=0$ radial envelope becomes
\begin{equation}
\psi_{0,m}(r)=\sqrt{\frac{2}{\pi\ell_{\mathrm{ho}}^{2}\Gamma(|m|+1)}}
\left(\frac{r}{\ell_{\mathrm{ho}}}\right)^{|m|}
e^{-r^{2}/(2\ell_{\mathrm{ho}}^{2})},
\end{equation}
where the power-law prefactor ensures proper normalization:
$$
\int_0^\infty |\psi_{0,m}(r)|^2 r\,dr = 1.
$$
This structure reflects the interplay between the centrifugal barrier and the torsion-induced harmonic trap, resulting in a suppression of the wavefunction near the origin for $|m| > 0$. As shown in Fig.~3, this results in a maximum of the current density at a finite radius $r_\text{max} \simeq \ell_{\mathrm{ho}}/\sqrt{2}$, rather than at $r=0$.

\paragraph*{Local electric current density.} Multiplying the probability current
$\mathbf{J}=(\hbar/m^{\ast})\Im(\Psi^{\ast}\nabla\Psi)$ by the charge $e$ gives \cite{sakurai}
\begin{equation}
\mathbf{j}_{xy}(r,\phi)
=\frac{e \hbar m}{m^{\ast}}\frac{|\Psi_{xy}(r,\phi)|^{2}}{r}
\hat{\boldsymbol\phi}\quad(\mathrm{Am^{-1}}).
\end{equation}
This expression highlights the linear dependence on the chirality index $m$ and the $1/r$ decay characteristic of azimuthal currents in cylindrical confinement.
\begin{figure}[t]
\centering
\includegraphics[width=\linewidth]{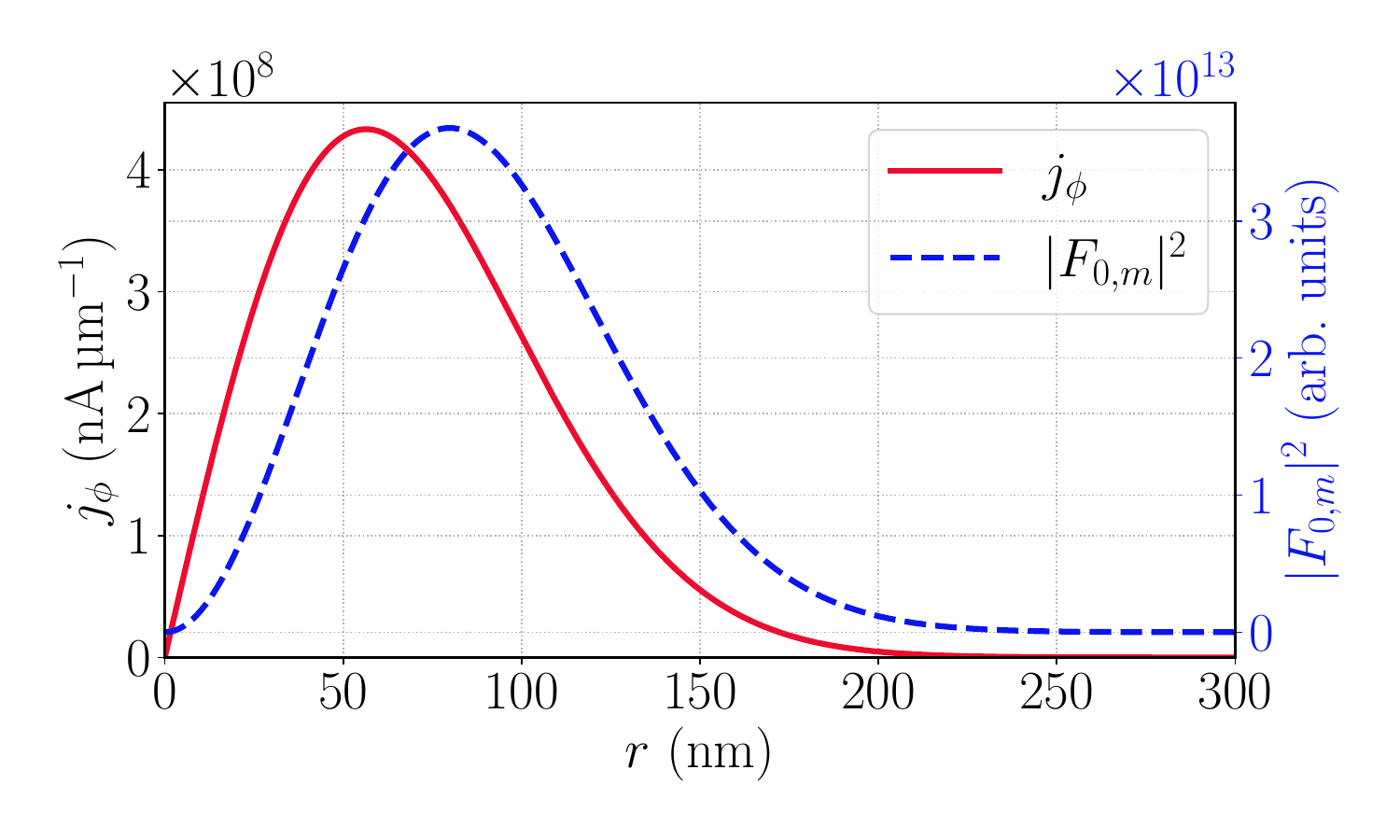}
\caption{\label{fig:jphi_profile} Radial dependence of the azimuthal persistent charge current $j_{\phi}(r)$ (solid red) for the ground branch $(n_r,m,n_z)=(0,+1,1)$ at torsion $\Omega=5\times10^{6}\,\mathrm{m}^{-1}$ and well length $L=100$\,nm. The left axis gives $j_{\phi}$ in $\mathrm{nA\,\mu m^{-1}}$, while the dashed blue curve (right axis) shows the corresponding transverse probability density $|F_{0,m}(r)|^{2}$ (arbitrary units). The current vanishes at the origin, peaks near $r\simeq\ell_{\mathrm{ho}}/\!\sqrt{2}\approx 45$\,nm, and decays exponentially for $r\gtrsim\ell_{\mathrm{ho}}$, in accordance with the analytic expression $j_{\phi}\propto re^{-r^{2}/\ell_{\mathrm{ho}}^{2}}$. The area under the green curve yields the total circulating current $I_{\mathrm{circ}}\simeq4\times10^{-8}$A.}
\end{figure}

Figure~\ref{fig:jphi_profile} quantifies the spatial structure of the persistent charge current discussed above. The current density $j_{\phi}(r)$ (green) is evaluated for the lowest radial quantum number $n_r=0$ and $m=+1$, using the analytic envelope derived from the harmonic confinement in Eq.~\eqref{eq:Veff_full}. Because $j_{\phi}\propto r^{-1}|F_{0,m}|^{2}$, the profile inherits the Gaussian factor $e^{-r^{2}/\ell_{\mathrm{ho}}^{2}}$ of the ground-state wavefunction but is further multiplied by a linear $r$ term that suppresses the current at $r=0$ and shifts the maximum to $r_{\mathrm{max}}\simeq\ell_{\mathrm{ho}}/\sqrt{2}$.

With the InAs parameters used throughout the paper ($\Omega=5\times10^{6}\,\mathrm{m}^{-1}$, $L=100$\,nm, $m^{\ast}=0.023\,m_{e}$), we obtain $\ell_{\mathrm{ho}}\approx64$\,nm and a peak current density of $j_{\phi}^{\mathrm{max}}\approx4.4\times10^{8}\,\mathrm{nA\,\mu m^{-1}}$. Integrating the curve gives a total circulating current $I_{\mathrm{circ}}\simeq4\times10^{-8}$A, in excellent agreement with the order-of-magnitude estimate of Sec.~\ref{sec:current}. This confirms that the geometry-induced torsion alone generates a measurable chiral current without the need for an external magnetic field.

\paragraph*{Numerical estimate for the InAs well.}
Using the parameters of Fig.~\ref{fig:Jphi} ($L=100$\,nm, $\Omega=5\times10^{6}\,\mathrm{m^{-1}}$, $m^{\ast}=0.023\,m_{e}$, $m=+1$) one finds $\ell_{\mathrm{ho}}\simeq64$\,nm and
\begin{equation}
j_{\phi}^{\max}\approx
\frac{e \hbar}{m^{\ast}\ell_{\mathrm{ho}}}
\frac{1}{\pi\ell_{\mathrm{ho}}^{2}}
\simeq8\times10^{-9}\ \mathrm{Am^{-1}},
\end{equation}
a value comfortably within the sensitivity of state-of-the-art nano-SQUID current detectors.
\begin{figure}[tbhp]
\centering
\includegraphics[width=0.94\columnwidth]{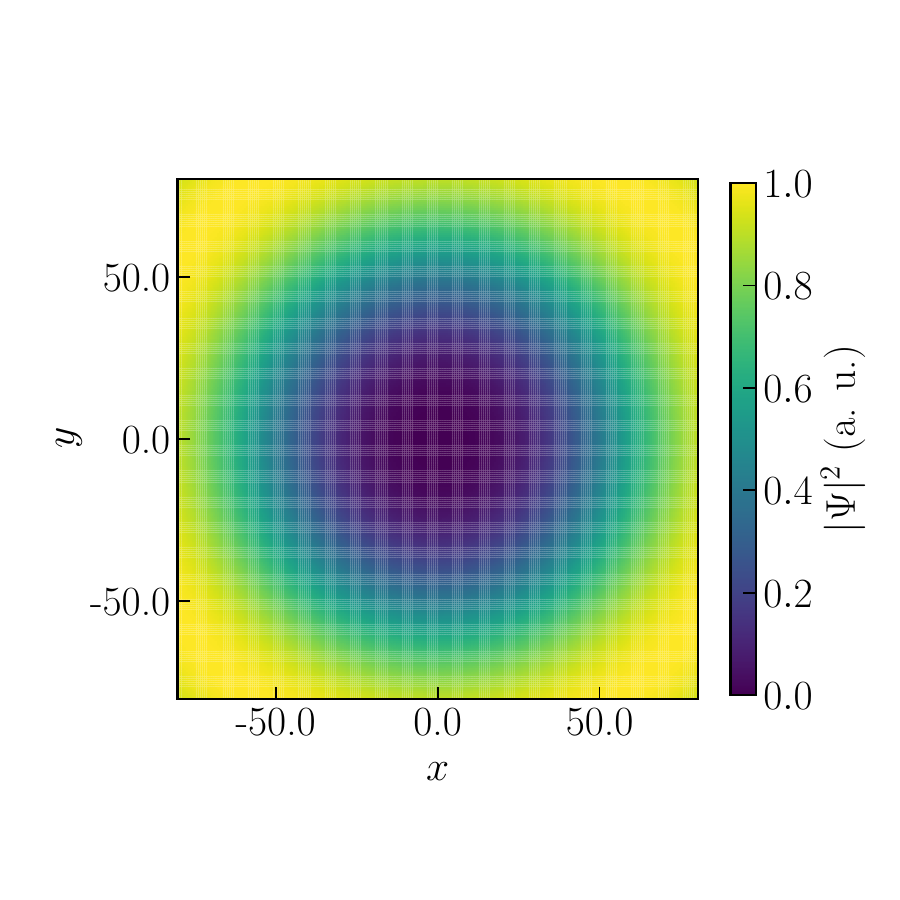}\vspace{-1.0cm}
\caption{\label{fig:density}
Normalised transverse probability density $|\Psi_{xy}(x,y)|^{2}$ for the ground state $(n_{r},m,n_{z})=(0,+1,1)$ at torsion $\Omega=5\times10^{6}\,\mathrm{m}^{-1}$. The radial extent is set by $\ell_{\mathrm{ho}}\simeq64$\,nm.}
\end{figure}
\begin{figure}[t]
\centering
\includegraphics[width=0.95\columnwidth]{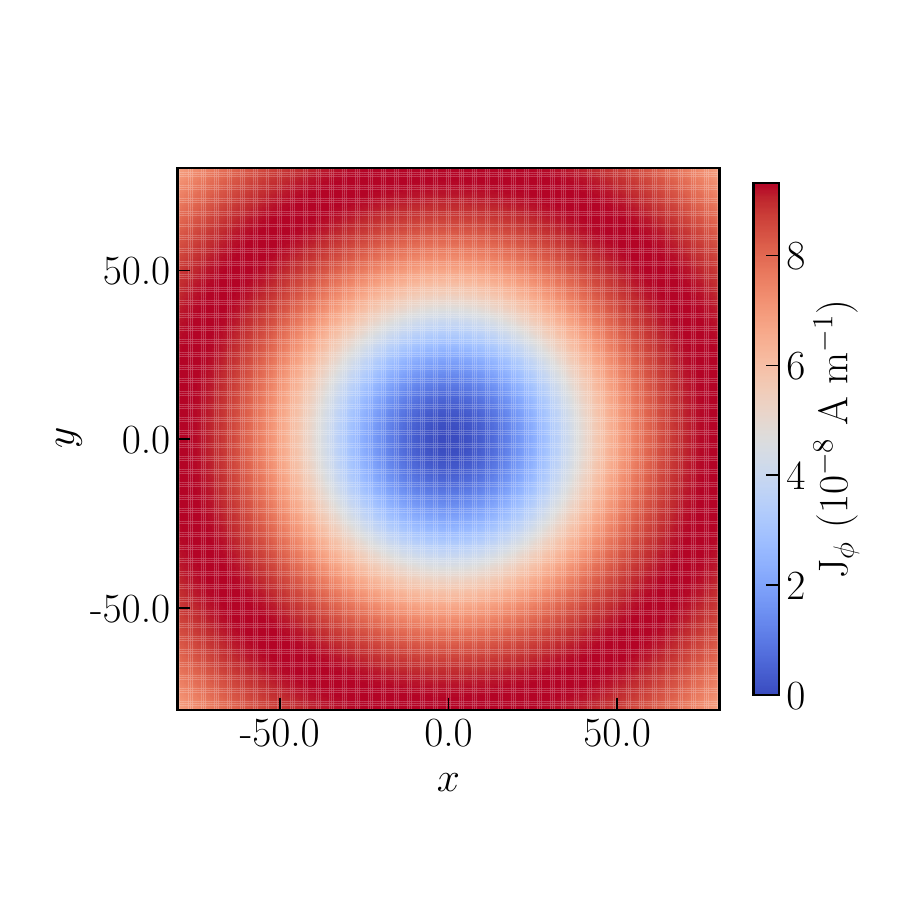}\vspace{-1.0cm}
\caption{\label{fig:Jphi}
Azimuthal electric current density $J_{\phi}(x,y)$ for the same state. The peak value $J_{\phi}^{\max}\approx8\times10^{-9}$ Am$^{-1}$ agrees with the analytic estimate.}
\end{figure}
\begin{figure}[ht]
\centering
\includegraphics[width=\linewidth]{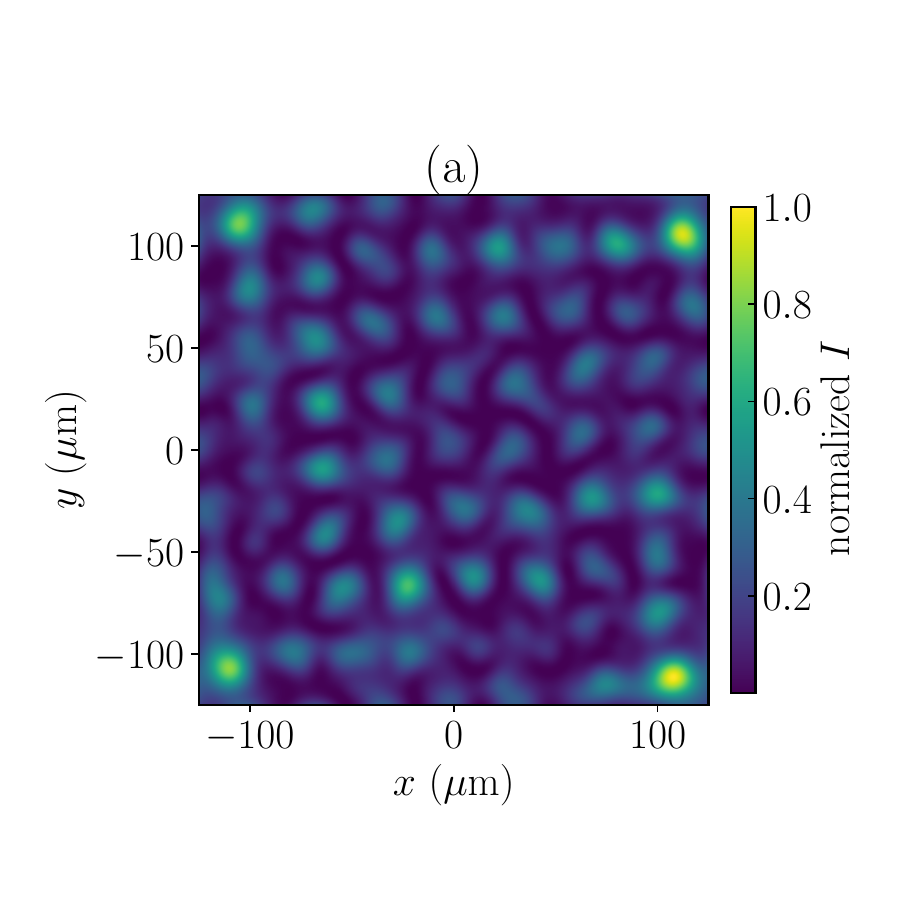}\vspace{-1.0cm}
\includegraphics[width=\linewidth]{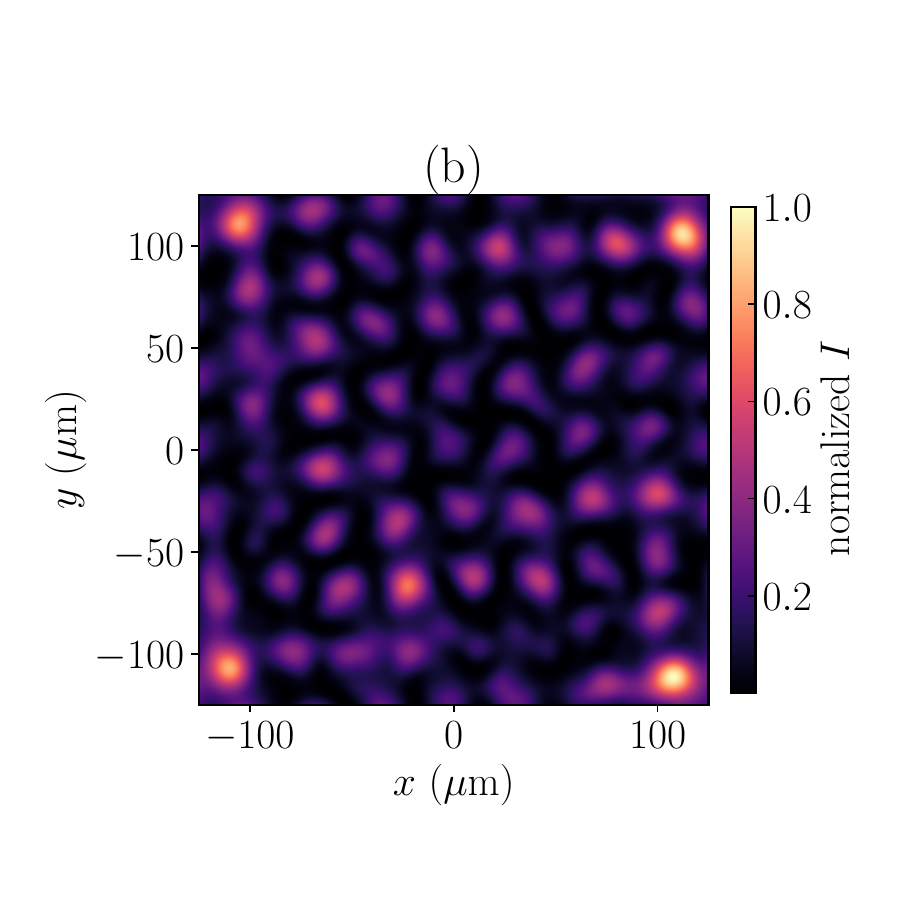}\vspace{-1.0cm}
\caption{\textbf{(a)}~Bulk Landau mode. Normalized transverse intensity $I(x,y)$ for the lowest Landau level in the helically twisted medium. The mode fills the interior in a nearly uniform pattern of Landau orbitals. 
\textbf{(b)}~Chiral edge mode. Normalized intensity of a topological boundary state, tightly localized along the edge and circulating in a single direction.}
\label{fig:modes}
\end{figure}

\begin{figure}[tbhp]
\centering
\includegraphics[width=\linewidth]{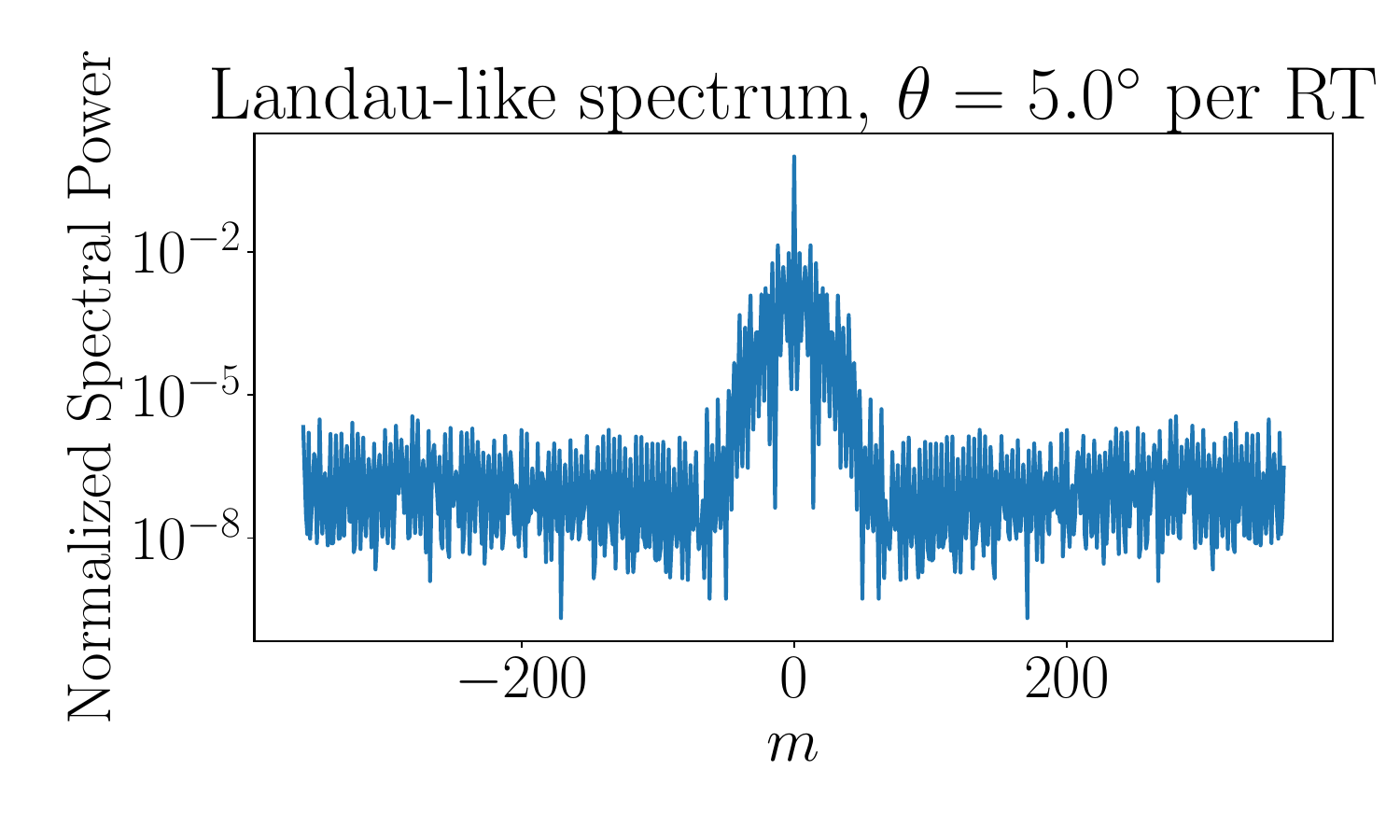}
\caption{Landau-like spectrum. Normalized power spectrum versus azimuthal index $m$ for the bulk mode shown in Fig.~\ref{fig:modes}(a). The sharp peak at $m=0$ and exponential suppression of higher-order components ($|m| > 0$) are hallmarks of Landau quantization. Here, the spectrum is obtained with a torsion-induced rotation rate of $\theta = 5.0^\circ$ per round trip (RT), mimicking the effect of a synthetic magnetic field. This geometrically induced splitting illustrates the emergence of discrete angular momentum states without the need for external fields or potentials.}
\label{fig:landau_spectrum}
\end{figure}

\section{Light propagation in a twisted helicoidal optical medium}
\label{sec:light}

We now investigate the effect of torsion on the paraxial propagation of a monochromatic light beam in a helicoidal optical structure. Our starting point is the scalar Helmholtz equation for a time-harmonic field in flat space \footnote{Standard optics derivations can be found in Saleh and Teich, Fundamentals of Photonics, 3rd ed. (Wiley, 2019).},
\begin{equation}
\nabla^2 \Psi + k_0^2 n_0^2 \Psi = 0,
\label{eq:helmholtz}
\end{equation}
where $\Psi(x, y, z)$ is the total complex field, $k_0 = 2\pi/\lambda_0$ is the vacuum wavenumber, and $n_0$ is the uniform background refractive index.

To describe beam-like solutions predominantly propagating along the $z$-axis, we employ the standard ansatz
\begin{equation}
\Psi(x, y, z) = \psi(x, y, z) e^{i k_0 n_0 z},
\label{eq:ansatz}
\end{equation}
where $\psi(x,y,z)$ is a slowly-varying envelope function. Substituting this into Eq.~\eqref{eq:helmholtz}, we have
\begin{equation}
\nabla_\perp^2 \psi + \frac{\partial^2 \psi}{\partial z^2}
       + 2i k_0 n_0 \frac{\partial \psi}{\partial z} = 0,
\end{equation}
where $\nabla_\perp^2 = \partial_x^2 + \partial_y^2$ defines the transverse Laplacian. Assuming that the envelope varies slowly with $z$, we apply the paraxial approximation by neglecting the term $\partial_z^2 \psi$, leading to
\begin{equation}
2i k_0 n_0 \frac{\partial \psi}{\partial z}+ 
\nabla_\perp^2 \psi \approx 0.
\end{equation}
Dividing both sides by $2k_0 n_0$, we finally obtain the paraxial wave equation
\begin{equation}
i \frac{\partial \psi}{\partial z}= -\frac{1}{2k_0 n_0} 
\nabla^2_\perp \psi,
\label{eq:paraxial_free}
\end{equation}
which is structurally analogous to the Schrödinger equation in quantum mechanics, where the propagation axis $z$ plays the role of time.

However, to faithfully describe the effect of torsion inspired by a helicoidal space-time geometry, we revisit the derivation from a curved-space formulation. Starting from a line element with helicoidal torsion (\ref{eq:metric_cart}) and assuming scalar wave propagation, the d'Alembertian operator leads to a modified Helmholtz equation that naturally incorporates torsional effects. Under the paraxial and slowly varying envelope approximations, the resulting evolution equation for the optical field envelope becomes
\begin{equation}
i \frac{\partial \psi}{\partial z}
   = -\frac{1}{2k_0 n_0} 
\nabla^2_\perp \psi
     + V_\text{tor}(x,y)\psi,
\label{eq:paraxial_torsion}
\end{equation}
in which the torsion-induced potential arises from the off-diagonal metric components:
\begin{equation}
V_\text{tor}(r, \phi)
   = -\frac{\Omega^2 r^2}{2k_0 n_0}
     \frac{\partial^2}{\partial \phi^2},
\label{eq:torsion_potential_cyl}
\end{equation}
where $\Omega$ denotes the torsion strength and $(r,\phi)$ are cylindrical coordinates.

Although this potential is most naturally written in cylindrical coordinates, we perform the numerical simulation on a Cartesian grid. Using the identity $\partial_\phi= -y \partial_x + x \partial_y$, the torsion potential becomes
\begin{equation}
V_\text{tor}(x, y) = -\frac{\Omega^2}{2k_0 n_0}(x^2 + y^2)
\bigl(-y \partial_x + x \partial_y\bigr)^2.
\label{eq:torsion_potential_cartesian}
\end{equation}
This retains the cylindrical symmetry while enabling efficient FFT-based propagation on a Cartesian mesh.

We use the standard split-step method, alternating (i) the linear diffraction term in Fourier space and (ii) the torsion potential in real space. The input beam is a fundamental Gaussian
\begin{equation}
\psi(x, y, 0)=e^{-\tfrac{x^{2}+y^{2}}{w_0^{2}}},
\quad w_0 = 9.0\,\mu\text{m},
\end{equation}
propagated over $N_z = 10000$ steps of $\Delta z = 1.0\,\mu$m, totalling $10$mm inside a glass-like medium with $n_0 = 1.45$ at $\lambda_0 = 1.55\,\mu$m. The resulting field $\psi(x,y,z_\text{final})$ encodes the torsion-induced redistribution of intensity and phase, as visualized below with a standard color scale for direct comparison.

\begin{figure*}[tbhp]
\centering
\includegraphics[scale=0.50]{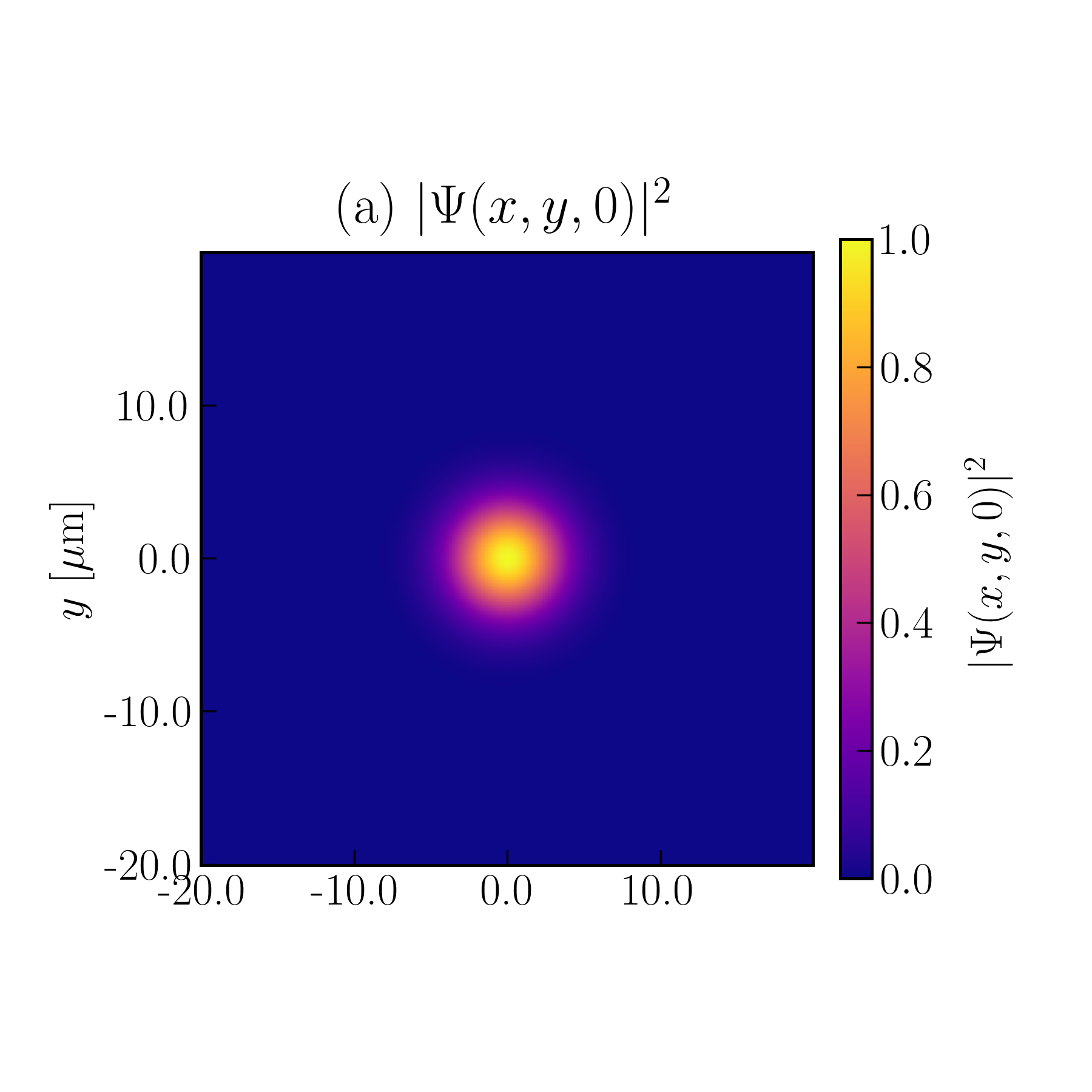}
\includegraphics[scale=0.50]{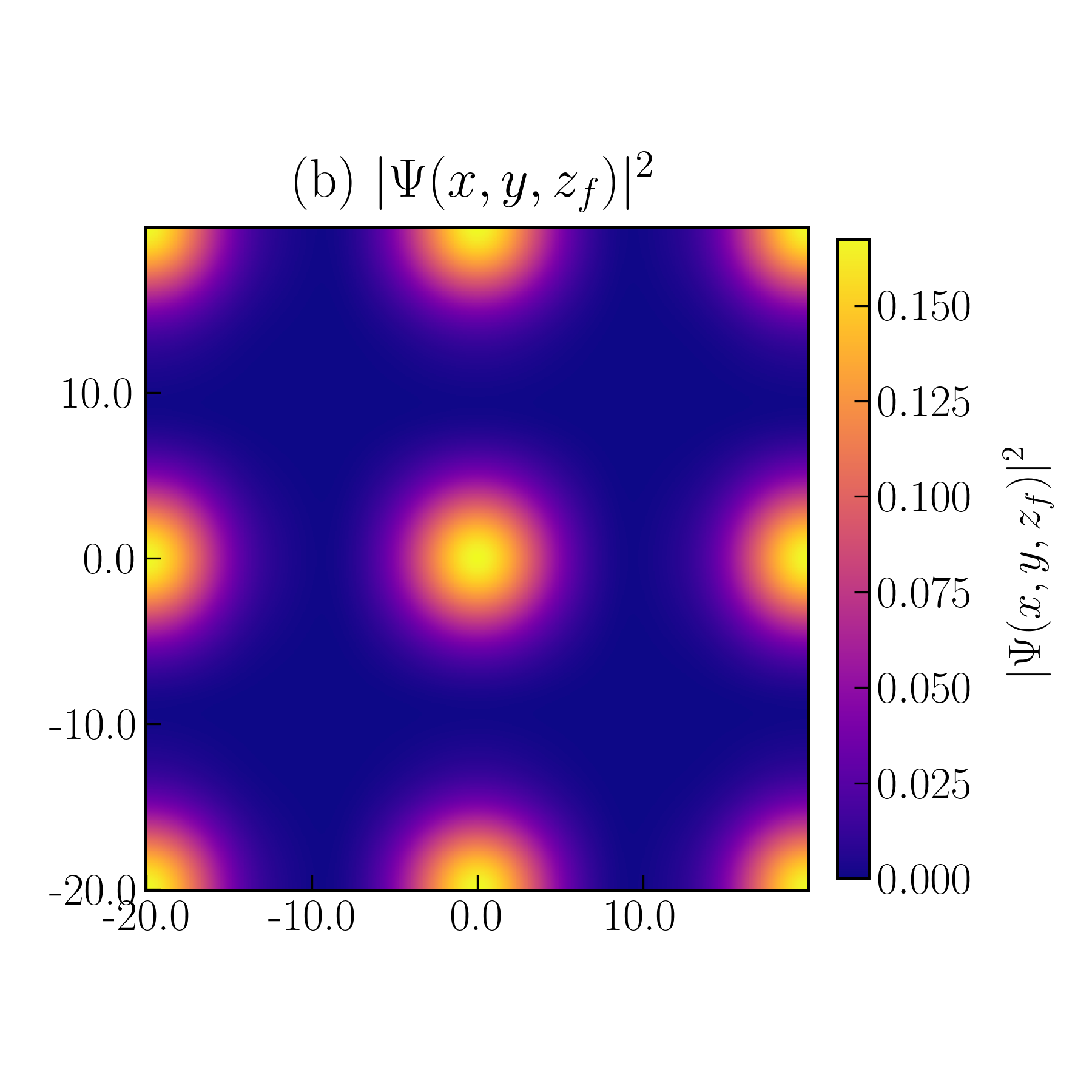}\vspace{-1.0cm}
\includegraphics[scale=0.50]{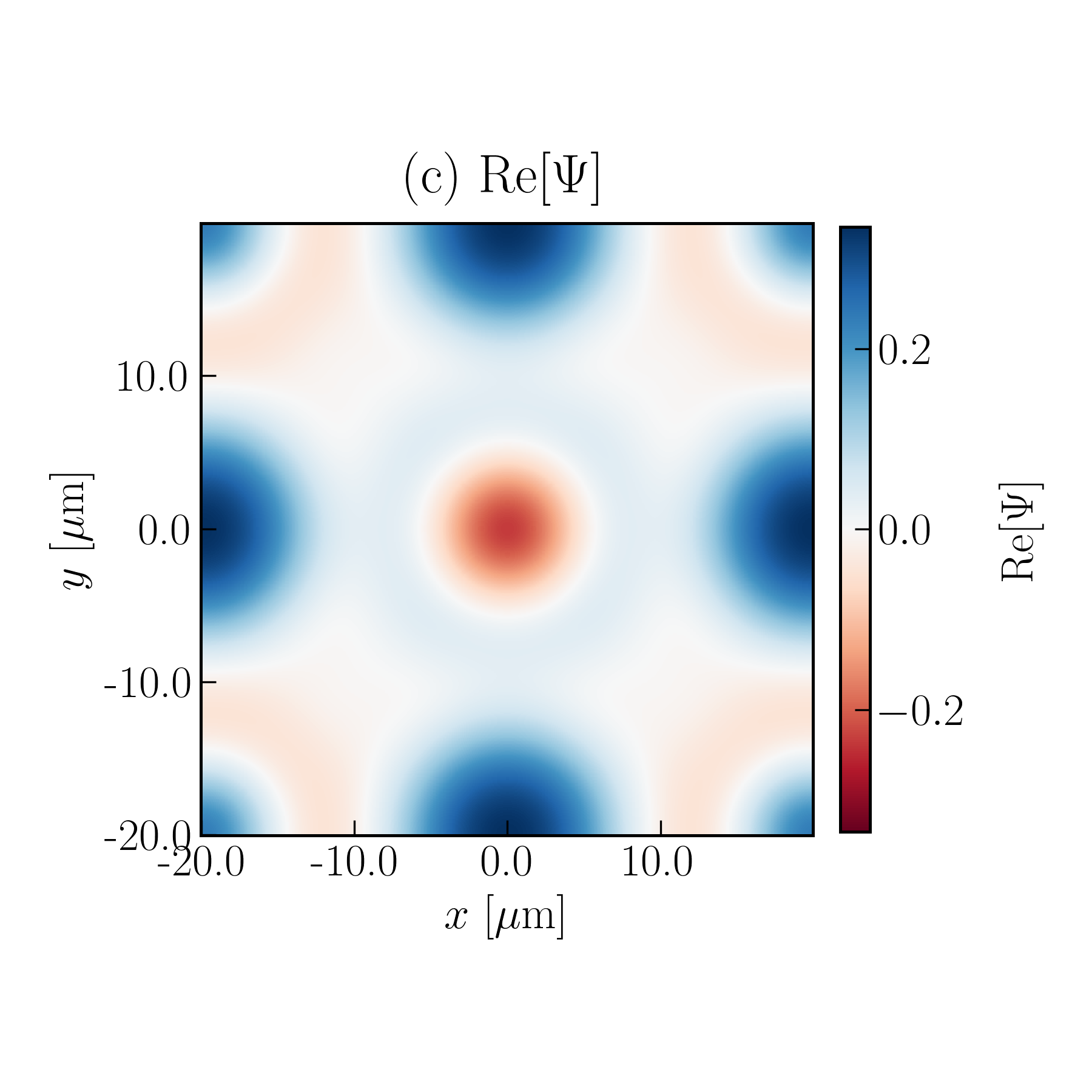}
\includegraphics[scale=0.50]{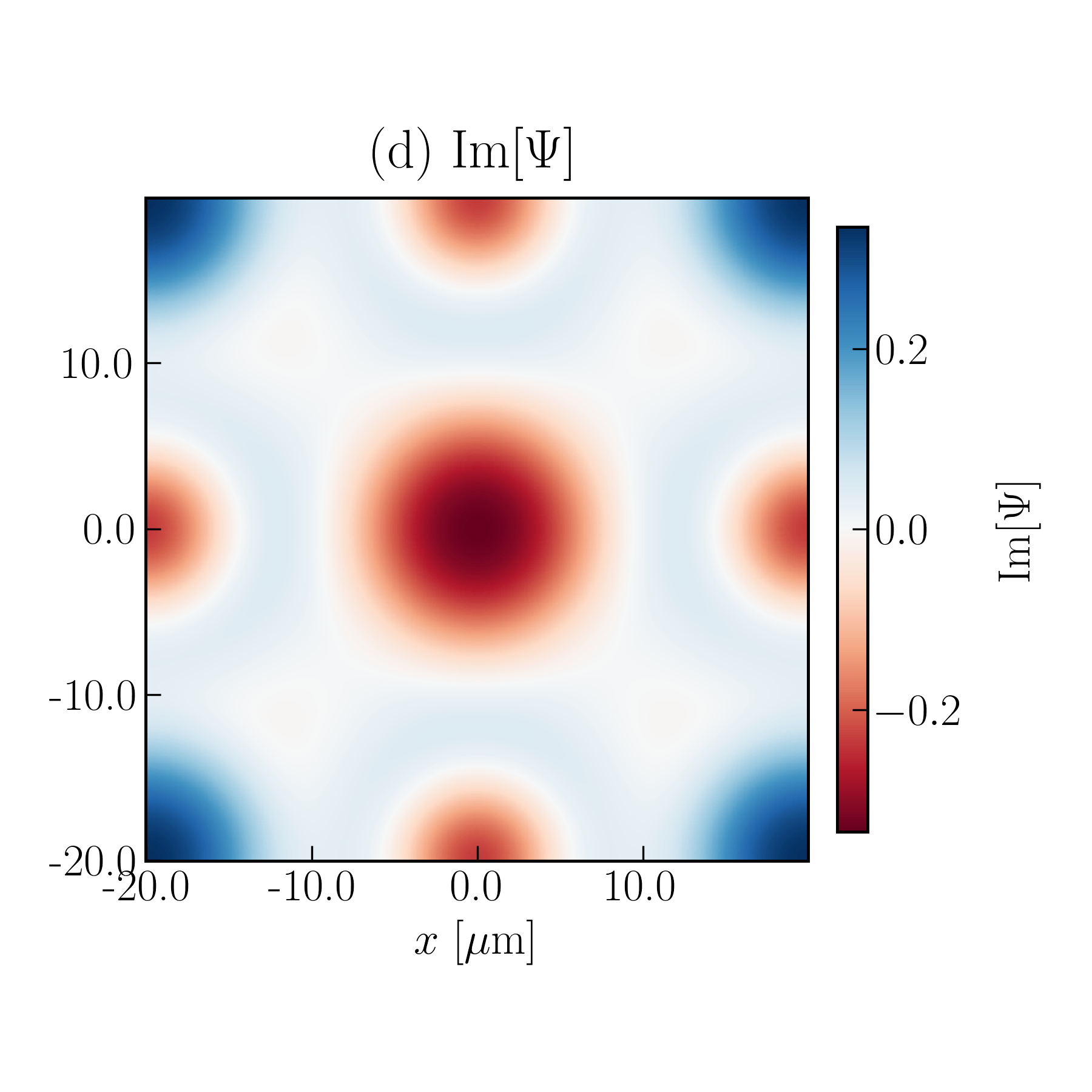}\vspace{-1.0cm}
\caption{Torsion-induced transformation of the optical field.
\textbf{(a)}~Initial Gaussian intensity profile at $z=0$. 
\textbf{(b)}~Final intensity after propagation through the helicoidal medium, showing a multi-lobed structure.
\textbf{(c)}~Real part and 
\textbf{(d)}~imaginary part of the output complex field. Together, they highlight the emergence of phase singularities and chiral phase fronts.}
\label{fig:helicoidal_beam}
\end{figure*}

\begin{figure*}[ht]
\centering
\includegraphics[scale=0.5]{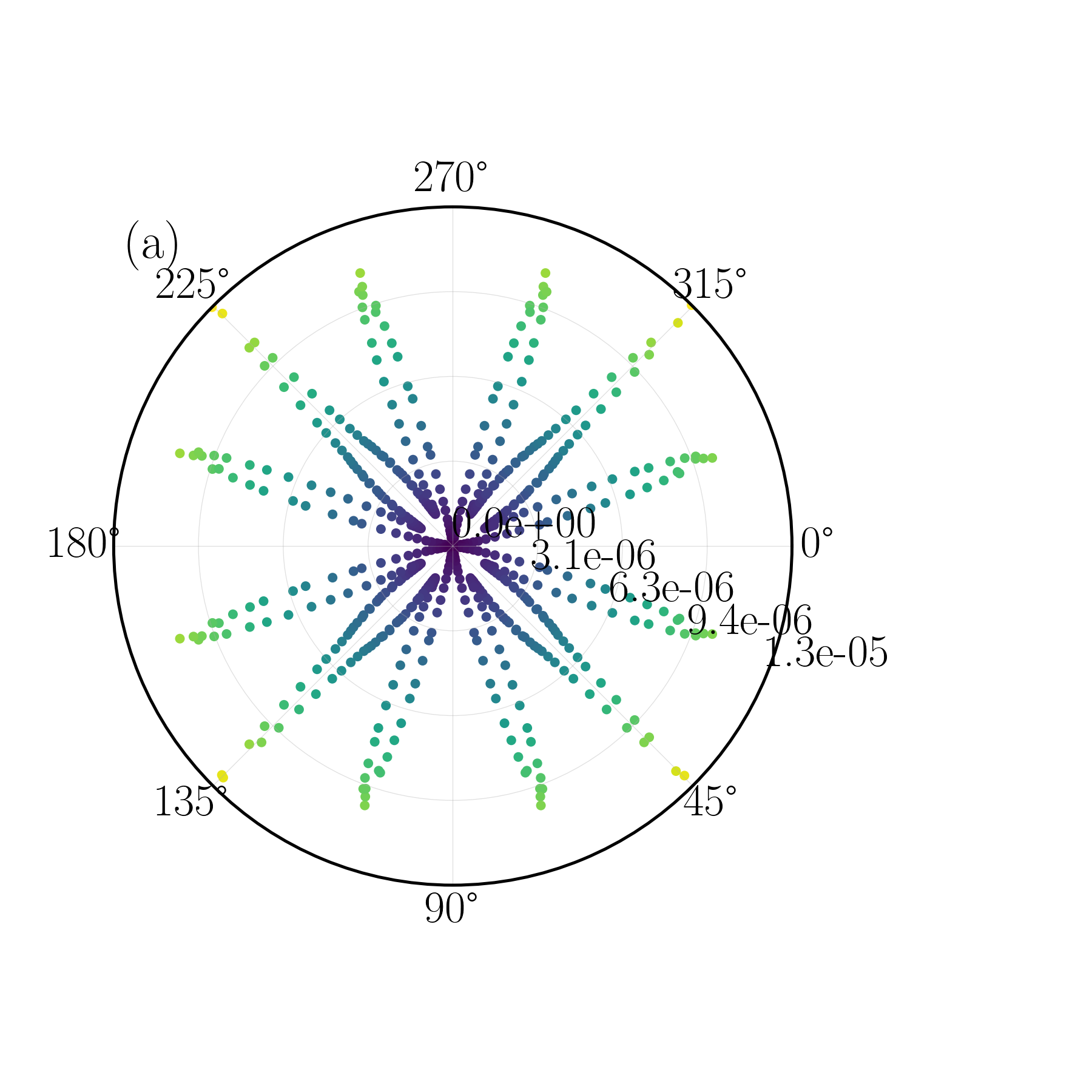}
\includegraphics[scale=0.5]{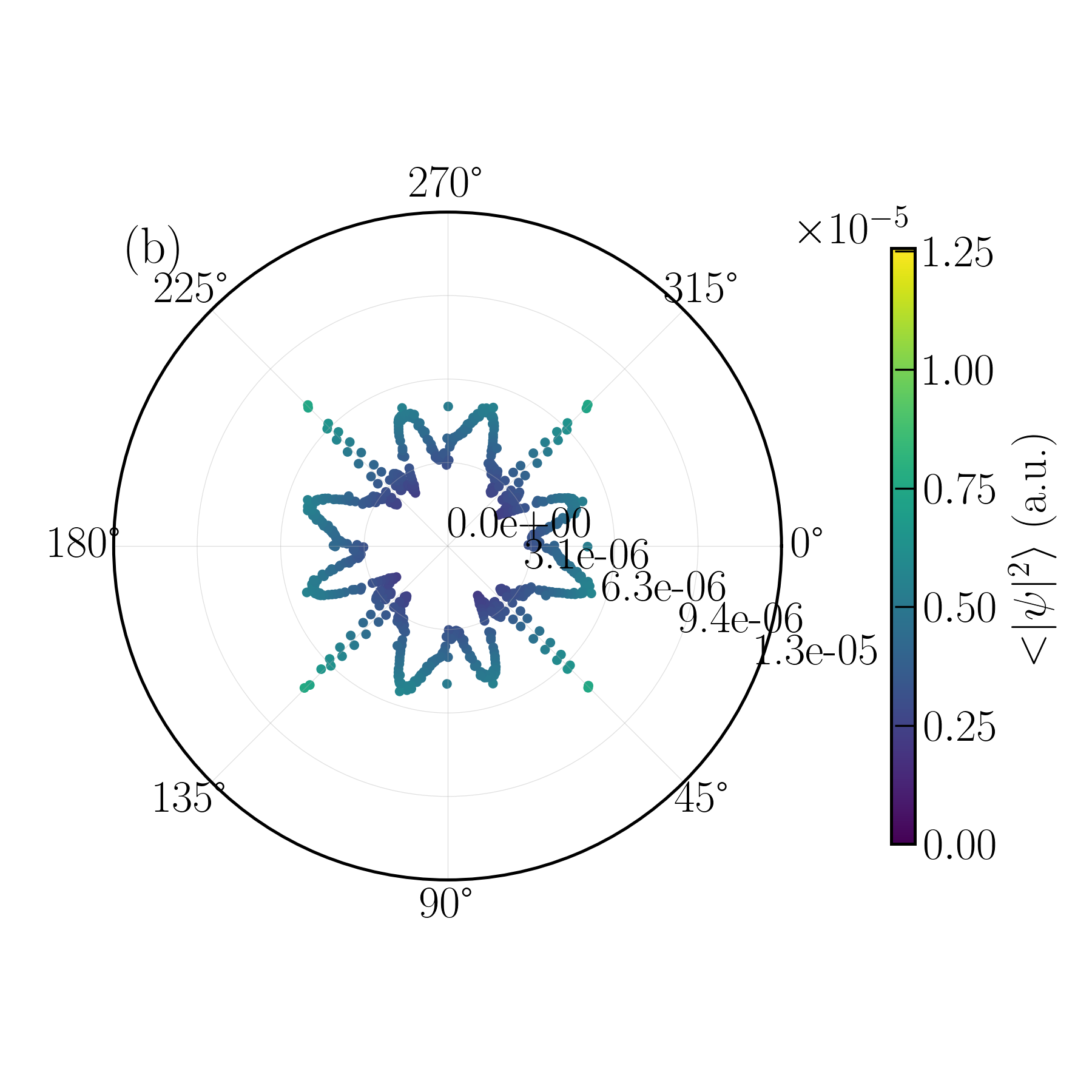}\vspace{-1.0cm}
\includegraphics[scale=0.5]{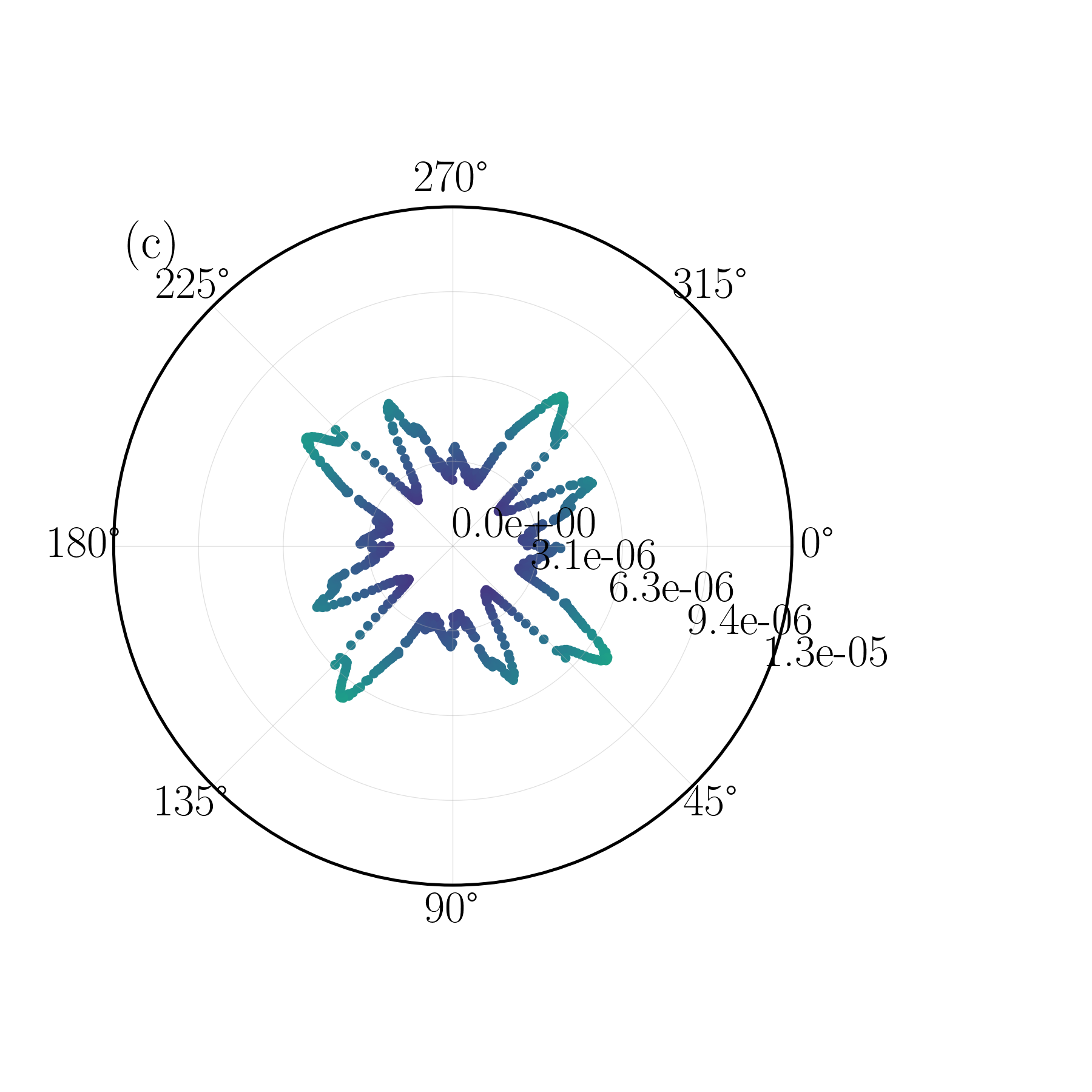}
\includegraphics[scale=0.5]{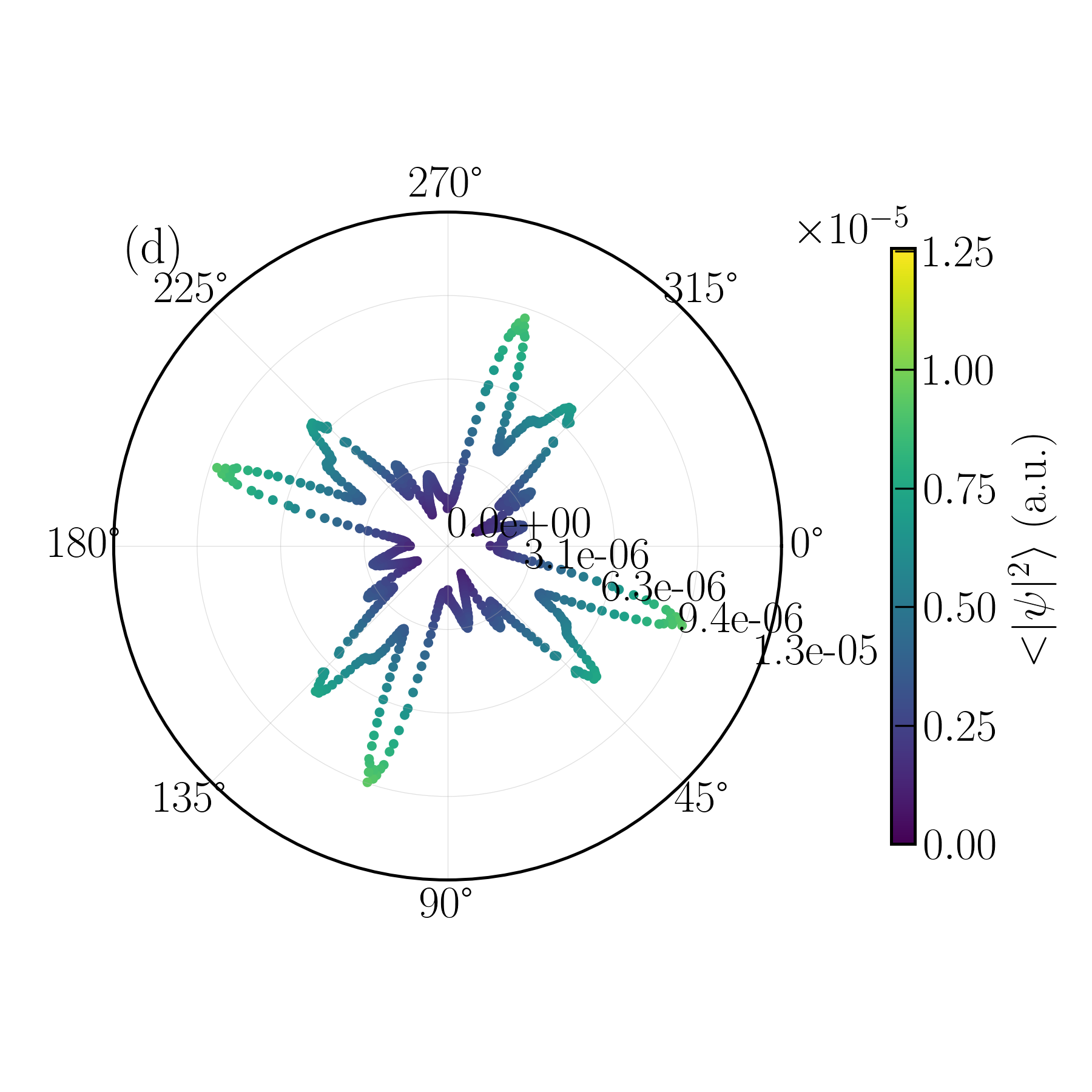}\vspace{-1.0cm}
\caption{Normalised polar intensity maps $\langle|\psi|^2\rangle(\phi)$ after 10 mm propagation inside the helicoidal medium. The angular coordinate $\phi$ is measured clockwise from the $x$-axis, and the radial coordinate is normalised to the global maximum intensity. All panels share the same colour scale. Increasing torsion drives a clear sequence of structural transformations: (a) an eightfold symmetric pattern at zero torsion ($\Omega = 0$), indicating no preferential direction; (b) uniform contraction under weak torsion, showing early signs of symmetry breaking; (c) a star-shaped alternation of long and short lobes at intermediate torsion, revealing azimuthal mode mixing; and (d) four dominant lobes once the $\pm m$ degeneracy is fully lifted, mimicking Landau-level quantization with discrete angular momentum states. This progression illustrates how geometric torsion alone can induce complex spatial structures and break chiral symmetry in classical optical fields.}
\label{fig:rosacea_grid}
\end{figure*}

\section{Optical realization: Landau-level physics in a helically twisted medium}
\label{sec:optical_landau}

In recent years, the creation of synthetic gauge fields for photons has opened the door to photonic analogues of electronic topological phenomena \cite{PRApp.2024.22.014047,NP.2024.20.1738,PRApp.2025.23.054080,photonics.2025.12.612}. By engineering a helically twisted refractive-index landscape, an effective magnetic field can be induced for light propagating in the transverse plane, resulting in Landau-level quantization and robust chiral edge modes \cite{PRB.2016.93.235315,PRA.2016.94.043611,PRB.2024.110.165421,PRB.2025.111.L140507}. Such effects promise new avenues for disorder-immune optical guiding, beam shaping, and on-chip topological lasers.

In our implementation, a paraxial Gaussian beam propagates through a twist-imprinted optical cavity (or waveguide array) whose local rotation rate $\Omega$ plays the role of a magnetic field. Under the paraxial and slowly varying envelope approximations, the torsion enters as a quadratic``vector-potential" term in the transverse Helmholtz equation, giving rise to discrete Landau levels and topologically protected boundary states.

\paragraph*{Bulk and edge modes.}
Figure~\ref{fig:modes} presents the transverse intensity patterns resulting from light propagation through the helically twisted medium. Panel~(a) shows the normalized intensity $I(x,y)$ of the bulk Landau mode, exhibiting a quasi-periodic array of peaks across the interior of the transverse plane ($|x|,|y|\lesssim100\,\mu$m), consistent with the highly degenerate lowest Landau level. This degeneracy is characteristic of systems under synthetic magnetic fields and has been previously observed in photonic lattices where geometric torsion mimics the effect of an external magnetic field~\cite{Rechtsman2013}. The spatial distribution of these peaks reflects the interplay between the harmonic trapping potential induced by the metric and the chiral coupling between azimuthal and axial motion. The geometric torsion-induced potential in our model finds parallels in optical systems where structured media emulate magnetic fields for light~\cite{JPCM.2008.20.125209}. In both cases, spatial twisting imprints a synthetic vector potential that leads to discrete angular momentum states and chiral edge modes.

Panel~(b) displays the corresponding chiral edge mode, where light is confined to the periphery ($|x|\approx100 \,\mu$m or $|y|\approx100 \,\mu$m) and propagates in a single direction determined by the sign of the twist $\Omega$. These edge states are topologically protected and robust against local perturbations, similar to those found in quantum Hall systems~\cite{RMP.2010.82.3045}. Their unidirectional nature makes them promising candidates for applications in non-reciprocal waveguides and topological photonics~\cite{RMP.2019.91.015006}. Notably, the absence of backscattering channels ensures minimal loss and high-fidelity signal transmission, which can be exploited in integrated optical circuits.

\paragraph*{Landau-level spectrum.}
Figure~\ref{fig:landau_spectrum} shows the normalized spectral power versus azimuthal index $m$ on a logarithmic scale. The sharp peak at $m=0$ and exponential suppression of sidebands confirm the formation of a strongly localized lowest Landau level, consistent with the bulk mode of Fig.~\ref{fig:modes}(a). This spectral profile is a direct consequence of the effective harmonic potential generated by the torsion parameter $\Omega$, as described in Eq.~(12) of the main text.

The localization of the lowest angular momentum state at $m=0$ mirrors the physics of Landau quantization in electronic systems, where the lowest energy state corresponds to vanishing orbital angular momentum~\cite{PRB.2011.83.205101,Nature.2017.8.50,PRL.2013.110.156403}. In our system, this localization arises purely from geometry, without any external gauge fields or magnetic fluxes. The exponential decay of higher-order components further supports the idea that the torsion-induced potential acts as a confining trap for the optical field, analogous to the magnetic confinement of charged particles in the quantum Hall regime.

Although the underlying equation governing this behavior, the paraxial Helmholtz equation, is classical in nature, its mathematical structure is isomorphic to the Schrödinger equation under a synthetic magnetic field. This allows us to rigorously define discrete Landau-like levels and chiral symmetry breaking in a photonic platform. The observed spectral structure thus provides compelling evidence that intrinsic geometric torsion alone can reproduce key features of Landau-level physics, even in the absence of quantum statistics or particle interactions.

This result establishes a clear correspondence between the photonic and quantum mechanical descriptions of Landau-level physics, reinforcing the central claim of the paper that intrinsic torsion can serve as a universal mechanism for inducing both confinement and chirality.

\paragraph*{Field components and phase structure.}
Figure~\ref{fig:helicoidal_beam} illustrates the evolution of the optical field under torsion. Panel~(a) shows the initial Gaussian intensity at $z=0$; panel~(b) shows the final intensity profile, featuring a multi-lobed structure that signals azimuthal mode mixing. This transformation demonstrates how the helicoidal metric induces coupling between different orbital angular momentum (OAM) states during propagation, leading to a redistribution of energy among various $m$-modes.

Panels~(c) and~(d) display the real and imaginary parts of the complex field at the output facet, revealing complementary phase patterns and the emergence of chiral phase fronts. These phase structures are consistent with the generation of persistent azimuthal currents, even in the absence of an external magnetic field, as predicted by the theoretical model in Section~III.

The presence of phase singularities, vortices where the phase is undefined, is particularly notable. These vortices correspond to regions of zero intensity and carry quantized phase windings, directly linking the optical behavior to the quantum mechanical description of persistent currents. The observed phase evolution provides strong numerical evidence for the existence of a geometrically induced Zeeman-like splitting between $+m$ and $-m$ states, mirroring the chiral effects discussed in Sections II-IV.

\paragraph*{Chiral confinement and mode evolution.}
Taken together, these results establish a photonic platform for simulating Landau-level physics: a degenerate bulk spectrum, robust chiral boundary states, and quantized angular-momentum modes. The torsion-induced confinement and spectral evolution seen in Fig.~\ref{fig:rosacea_grid} further reinforce the correspondence with quantum mechanical behaviour discussed in Sections~\ref{sec:geometry}-\ref{sec:current}.

As shown in the polar intensity maps $\langle|\psi|^2\rangle(\phi)$, increasing torsion $\Omega$ drives a sequence of structural transitions: from an eightfold symmetric pattern at $\Omega=0$, through a uniform contraction under weak torsion, to a star-shaped alternation of long and short lobes at intermediate torsion, and finally to four dominant lobes once the $\pm m$ degeneracy is fully lifted. This progression clearly demonstrates how torsion progressively breaks the rotational symmetry of the system, lifting the degeneracy between positive and negative angular momentum states.

These observations align closely with the analytic predictions derived from the Schrödinger equation in the quantum mechanical framework. In particular, the linear dependence of the chiral splitting on both $\Omega$ and $n_z$ [Eq.~(17)] is mirrored in the photonic domain through the torsion-dependent redistribution of intensity among azimuthal modes. This correspondence confirms the universality of the geometric effects introduced by the helicoidal metric, applicable not only to matter waves but also to classical electromagnetic waves.

The unified quantum-optical description highlights that both phenomena, chiral splitting and geometric confinement, are rooted in the same underlying metric structure, offering a new paradigm for engineering synthetic gauge fields and topological states using pure geometry.

\section{Experimental perspectives}
\label{sec:exp}

The predicted chiral splitting ($\sim 0.5\,\mathrm{meV}$ for typical InAs parameters with $L = 100\,\mathrm{nm}$ and $\Omega = 5 \times 10^6\,\mathrm{m^{-1}}$) opens several concrete experimental pathways, particularly in cold-atom, photonic, and strain-engineered platforms. This energy scale is not only within the resolution of current spectroscopic techniques but also comparable to other quantum effects such as spin-orbit coupling and Zeeman splittings in semiconductor heterostructures. As such, it represents an accessible regime for testing purely geometric quantum phenomena without the need for external magnetic fields or electrostatic gates.

Ultracold atoms trapped in tailored optical potentials constitute an ideal platform for realising our mesoscopic helicoidal quantum well. Optical painting with spatial light modulators (SLMs) can imprint helicoidal confinement potentials onto Bose or Fermi gases~\cite{Henderson2009}. A rotating anisotropic laser trap with radial and azimuthal modulation directly enforces the metric component $dz + \Omega r^{2}d\phi$, effectively encoding the torsion into the geometry experienced by the atomic cloud. Synthetic gauge fields generated by Raman coupling provide an additional handle to tune $\Omega$ dynamically, enabling precise control over the chirality of the system. Typical trap frequencies (a few kHz) match the chiral splitting scale, allowing detection by radio-frequency spectroscopy, direct \emph{in situ} absorption imaging, or time-of-flight interferometry~\cite{Ha2015}.

Femtosecond-laser-written photonic lattices offer a second route to probe the helicoidal geometry. By varying the writing speed and transverse modulation, one can inscribe waveguide arrays that follow helical trajectories with precisely controlled pitch and radius~\cite{Szameit2010}. Such structures implement the same Laplace-Beltrami operator that governs the quantum well, and torsion-induced mode splitting manifests as asymmetric intensity profiles and measurable shifts in the guided-mode spectrum. These features are observable using standard near-field microscopy or Fourier-space imaging techniques~\cite{Rechtsman2013}. The photonic platform also enables the direct visualization of bulk Landau-level analogues and topologically protected edge modes, providing a versatile testbed for exploring quantum-like phenomena in classical optics.

Rolled-up semiconductor heterostructures (e.g., InAs/GaAs nanomembranes) provide a solid-state incarnation of the helicoidal quantum well~\cite{Prinz2000}. Selective under-etching, combined with built-in strain gradients, produces tubes or helices with radii as small as hundreds of nanometres and torsion rates exceeding $10^{7}\,\mathrm{m^{-1}}$. Low-temperature transport measurements would detect the chiral splitting via magnetoconductance oscillations or weak-localisation signatures. These systems are particularly promising for applications in non-reciprocal electronics and chiral filters, where the absence of backscattering channels ensures robust unidirectional propagation of charge carriers.
\begin{table}[h]
    \centering
    \begin{tabular}{l c c c}
    \hline
    \textbf{} & \textbf{Cold atoms} & \textbf{Photonics} & \textbf{Nanorolls} \\
    \hline
    Radius $R$ & 30–50 $\mu$m & 20–50 $\mu$m & 0.2–0.5 $\mu$m \\
    Length $L$ & 0.5–1 mm & 5–20 mm & 0.1–2 $\mu$m \\
    Torsion $\Omega$ & $10^5$–$10^6$ m$^{-1}$ & $10^4$–$10^5$ m$^{-1}$ & $10^6$–$10^7$ m$^{-1}$ \\
    Detection & RF spec. & Near-field & Magnetotransp. \\
    \hline
    \end{tabular}
    \caption{Typical experimental parameters for three candidate platforms able to realize the mesoscopic helicoidal quantum well. Each column represents a key physical quantity: (i) \textbf{Radius ($R$)}: The transverse radius of the structure or beam confinement region; (ii) \textbf{Length ($L$)}: The total axial length of the system, corresponding to the quantization length in the $z$-direction; (iii) \textbf{Torsion ($\Omega$)}: The torsion strength, defined as the angular twist per unit length along the $z$-axis (in units of m$^{-1}$); (iv) \textbf{Detection}: The primary method used to observe the predicted geometric effects. For example: \emph{radio-frequency spectroscopy} (cold atoms), which measures energy level transitions; \emph{near-field imaging} (photonic systems), which visualizes optical intensity patterns close to the waveguide surface; and \emph{magnetotransport measurements} (nanorolls), which detect changes in electrical resistance due to chiral current splitting. These values illustrate that the required geometric parameters are within reach of current experimental capabilities across all three platforms.}
    \label{tab:experimental_parameters}
\end{table}

Table \ref{tab:experimental_parameters} summarises the typical parameter ranges achievable in these three distinct experimental platforms. Each operates within different length and torsion scales, yet all are capable of accessing the regime where geometric torsion-induced effects dominate over conventional potential landscapes. Notably, semiconductor nanorolls achieve the highest torsion rates ($10^6 - 10^7\,\text{m}^{-1}$), making them particularly suitable for probing sub-meV energy splittings. Meanwhile, cold-atom systems offer high tunability of $\Omega$ via external fields or modulation protocols. These comparisons underscore the broad applicability of the helicoidal geometry as a universal synthetic gauge field across multiple disciplines.

Our analysis neglects spin-orbit coupling and many-body interactions. In InAs, the Rashba term is of the same order as the geometric splitting for gate fields $\lesssim10^{5}\,\mathrm{Vm^{-1}}$; stronger electrostatic fields would require a combined treatment. In cold-atom realisations, three-body losses may limit coherence for densities above $10^{14}\,\mathrm{cm^{-3}}$. Addressing these effects forms a natural extension of the present work.

Looking forward, the interplay between intrinsic torsion and many-body physics presents a rich area for future exploration. For instance, helicoidal Luttinger liquids or interacting bosons in twisted optical traps could lead to novel fractionalized phases driven purely by geometry. On the photonic side, coupling multiple helicoidal waveguides into synthetic lattices offers a pathway to all-optical non-reciprocity and chiral quantum devices. Ultimately, this work demonstrates how geometry itself, through torsion, can act as a fundamental source of quantum effects, opening new avenues for both basic studies and technological applications.

\section{Conclusions}
\label{sec:conclusions}

We have introduced and analytically solved a genuinely geometric quantum well whose confinement, degeneracy lifting, and persistent currents stem exclusively from the intrinsic torsion of a helicoidal metric. Starting from the three-dimensional Heisenberg-group line element, we derived an effective radial Schrödinger equation whose potential contains: (i) the usual centrifugal barrier (with the Langer correction), (ii) a geometry-generated harmonic trap, and (iii) a chiral Zeeman shift proportional to $m$ and $k_z$. The resulting spectrum is fully analytic and reveals a tunable chiral splitting that extends to the sub-millielectronvolt range for realistic semiconductor parameters.

Complementing the quantum analysis, a scalar-paraxial study demonstrated that an optical beam propagating in a synthetic helicoidal medium obeys an identical equation with the same torsion-induced potential. Numerical split-step FFT simulations confirmed the expected regimes: eightfold symmetry at $\Omega = 0$, star-like patterns at intermediate torsion, and four dominant lobes once the $\pm m$ degeneracy is lifted.

Our approach builds on the geometric theory of defects originally developed for elastic media~\cite{JPCM.2008.20.125209}, but extends it to cold atoms, photonics, and semiconductor nanostructures. This work thus bridges the physics of torsion-induced quantisation in solids with emerging synthetic gauge fields in quantum technologies.

Three independent platforms can probe these predictions: (i) Cold atoms in SLM-painted helical traps, where radio-frequency spectroscopy can resolve the chiral splitting; (ii) Twisted photonic waveguides written by femtosecond laser-writing, already capable of sustaining the required twist rates and directly imaging mode profiles; and (iii) Strain-engineered nanorolls, in which magnetoconductance oscillations should expose the geometry-induced splitting on solid-state charge carriers. In all cases, the intrinsic geometric energy scale competes with, or even surpasses, state-of-the-art external gate potentials, opening a new design paradigm where geometry itself becomes the confining field.

Beyond the single-particle regime, many-body realizations, ranging from chiral Luttinger liquids on helicoidal edges to interacting bosons in twisted optical traps, promise rich physics, including torsion-driven fractionalization and topological pumping without the need for external gauge fields. On the photonic side, coupling several helicoidal waveguides into lattices suggests an all-geometric route to non-reciprocal light transport and chiral quantum optical devices. We therefore anticipate that the mesoscopic helicoidal quantum well introduced here will serve as a versatile playground for both fundamental tests of geometry-induced quantum phenomena and practical implementations of chiral transport in next-generation quantum technologies.

\section*{Acknowledgments}

This work was supported by CAPES (Finance Code 001), CNPq (Grant 306308/2022-3), and FAPEMA (Grants UNIVERSAL-06395/22 and APP-12256/22).

\bibliographystyle{apsrev4-2}
\input{Article_V3.bbl}
\end{document}

%% file: Article_V3.bbl
%

%% file: Article_V3.bbl
\begin{thebibliography}{38}%
\makeatletter
\providecommand \@ifxundefined [1]{%
 \@ifx{#1\undefined}
}%
\providecommand \@ifnum [1]{%
 \ifnum #1\expandafter \@firstoftwo
 \else \expandafter \@secondoftwo
 \fi
}%
\providecommand \@ifx [1]{%
 \ifx #1\expandafter \@firstoftwo
 \else \expandafter \@secondoftwo
 \fi
}%
\providecommand \natexlab [1]{#1}%
\providecommand \enquote  [1]{``#1''}%
\providecommand \bibnamefont  [1]{#1}%
\providecommand \bibfnamefont [1]{#1}%
\providecommand \citenamefont [1]{#1}%
\providecommand \href@noop [0]{\@secondoftwo}%
\providecommand \href [0]{\begingroup \@sanitize@url \@href}%
\providecommand \@href[1]{\@@startlink{#1}\@@href}%
\providecommand \@@href[1]{\endgroup#1\@@endlink}%
\providecommand \@sanitize@url [0]{\catcode `\\12\catcode `\$12\catcode
  `\&12\catcode `\#12\catcode `\^12\catcode `\_12\catcode `\%12\relax}%
\providecommand \@@startlink[1]{}%
\providecommand \@@endlink[0]{}%
\providecommand \url  [0]{\begingroup\@sanitize@url \@url }%
\providecommand \@url [1]{\endgroup\@href {#1}{\urlprefix }}%
\providecommand \urlprefix  [0]{URL }%
\providecommand \Eprint [0]{\href }%
\providecommand \doibase [0]{https://doi.org/}%
\providecommand \selectlanguage [0]{\@gobble}%
\providecommand \bibinfo  [0]{\@secondoftwo}%
\providecommand \bibfield  [0]{\@secondoftwo}%
\providecommand \translation [1]{[#1]}%
\providecommand \BibitemOpen [0]{}%
\providecommand \bibitemStop [0]{}%
\providecommand \bibitemNoStop [0]{.\EOS\space}%
\providecommand \EOS [0]{\spacefactor3000\relax}%
\providecommand \BibitemShut  [1]{\csname bibitem#1\endcsname}%
\let\auto@bib@innerbib\@empty
\bibitem [{\citenamefont {Nutbourne}\ and\ \citenamefont
  {Martin}(1988)}]{nutbourne1988differential}%
  \BibitemOpen
  \bibfield  {author} {\bibinfo {author} {\bibfnamefont {A.}~\bibnamefont
  {Nutbourne}}\ and\ \bibinfo {author} {\bibfnamefont {R.}~\bibnamefont
  {Martin}},\ }\href@noop {} {\emph {\bibinfo {title} {Differential Geometry
  Applied to Curve and Surface Design: Foundations}}},\ Differential Geometry
  Applied to Curve and Surface Design\ (\bibinfo  {publisher} {E. Horwood},\
  \bibinfo {year} {1988})\BibitemShut {NoStop}%
\bibitem [{\citenamefont {Lieberman}\ and\ \citenamefont
  {Lichtenberg}(2005)}]{liewberman2005principles}%
  \BibitemOpen
  \bibfield  {author} {\bibinfo {author} {\bibfnamefont {M.~A.}\ \bibnamefont
  {Lieberman}}\ and\ \bibinfo {author} {\bibfnamefont {A.~J.}\ \bibnamefont
  {Lichtenberg}},\ }\href@noop {} {\emph {\bibinfo {title} {Principles of
  Plasma Discharges and Materials Processing}}},\ \bibinfo {edition} {2nd}\
  ed.\ (\bibinfo  {publisher} {Wiley-Interscience},\ \bibinfo {address}
  {Hoboken, NJ},\ \bibinfo {year} {2005})\BibitemShut {NoStop}%
\bibitem [{\citenamefont {Nahmad-Achar}(2018)}]{book.Nahmad.2018}%
  \BibitemOpen
  \bibfield  {author} {\bibinfo {author} {\bibfnamefont {E.}~\bibnamefont
  {Nahmad-Achar}},\ }in\ \href {https://doi.org/10.1088/2053-2563/aadf65ch14}
  {\emph {\bibinfo {booktitle} {Differential Topology and Geometry with
  Applications to Physics}}},\ \bibinfo {series and number} {2053-2563}\
  (\bibinfo  {publisher} {IOP Publishing},\ \bibinfo {year} {2018})\ pp.\
  \bibinfo {pages} {14--1 to 14--11}\BibitemShut {NoStop}%
\bibitem [{\citenamefont {Li}\ \emph {et~al.}(2025)\citenamefont {Li},
  \citenamefont {Mehdi}, \citenamefont {Mehdi}, \citenamefont {Hussain},
  \citenamefont {Guo}, \citenamefont {Shi}, \citenamefont {Ali}, \citenamefont
  {Mehdi}, \citenamefont {Zhu}, \citenamefont {Ghaffar},\ and\ \citenamefont
  {Dhomeja}}]{OC.2025.577.131386}%
  \BibitemOpen
  \bibfield  {author} {\bibinfo {author} {\bibfnamefont {Y.}~\bibnamefont
  {Li}}, \bibinfo {author} {\bibfnamefont {I.}~\bibnamefont {Mehdi}}, \bibinfo
  {author} {\bibfnamefont {M.}~\bibnamefont {Mehdi}}, \bibinfo {author}
  {\bibfnamefont {S.}~\bibnamefont {Hussain}}, \bibinfo {author} {\bibfnamefont
  {J.}~\bibnamefont {Guo}}, \bibinfo {author} {\bibfnamefont {J.}~\bibnamefont
  {Shi}}, \bibinfo {author} {\bibfnamefont {S.}~\bibnamefont {Ali}}, \bibinfo
  {author} {\bibfnamefont {R.}~\bibnamefont {Mehdi}}, \bibinfo {author}
  {\bibfnamefont {S.}~\bibnamefont {Zhu}}, \bibinfo {author} {\bibfnamefont
  {A.}~\bibnamefont {Ghaffar}},\ and\ \bibinfo {author} {\bibfnamefont {L.~D.}\
  \bibnamefont {Dhomeja}},\ }\href
  {https://doi.org/https://doi.org/10.1016/j.optcom.2024.131386} {\bibfield
  {journal} {\bibinfo  {journal} {Optics Communications}\ }\textbf {\bibinfo
  {volume} {577}},\ \bibinfo {pages} {131386} (\bibinfo {year}
  {2025})}\BibitemShut {NoStop}%
\bibitem [{\citenamefont {Zhang}\ \emph {et~al.}(2023)\citenamefont {Zhang},
  \citenamefont {Li}, \citenamefont {Huang}, \citenamefont {Zhou},\ and\
  \citenamefont {Liang}}]{photonics.2023.10.1025}%
  \BibitemOpen
  \bibfield  {author} {\bibinfo {author} {\bibfnamefont {Y.}~\bibnamefont
  {Zhang}}, \bibinfo {author} {\bibfnamefont {B.}~\bibnamefont {Li}}, \bibinfo
  {author} {\bibfnamefont {T.}~\bibnamefont {Huang}}, \bibinfo {author}
  {\bibfnamefont {G.}~\bibnamefont {Zhou}},\ and\ \bibinfo {author}
  {\bibfnamefont {Y.}~\bibnamefont {Liang}},\ }\bibfield  {journal} {\bibinfo
  {journal} {Photonics}\ }\textbf {\bibinfo {volume} {10}},\ \href
  {https://doi.org/10.3390/photonics10091025} {10.3390/photonics10091025}
  (\bibinfo {year} {2023})\BibitemShut {NoStop}%
\bibitem [{\citenamefont {Schmidt}\ \emph {et~al.}(2020)\citenamefont
  {Schmidt}, \citenamefont {Weckesser}, \citenamefont {Thielemann},
  \citenamefont {Schaetz},\ and\ \citenamefont {Karpa}}]{PRL.2020.124.053402}%
  \BibitemOpen
  \bibfield  {author} {\bibinfo {author} {\bibfnamefont {J.}~\bibnamefont
  {Schmidt}}, \bibinfo {author} {\bibfnamefont {P.}~\bibnamefont {Weckesser}},
  \bibinfo {author} {\bibfnamefont {F.}~\bibnamefont {Thielemann}}, \bibinfo
  {author} {\bibfnamefont {T.}~\bibnamefont {Schaetz}},\ and\ \bibinfo {author}
  {\bibfnamefont {L.}~\bibnamefont {Karpa}},\ }\href
  {https://doi.org/10.1103/PhysRevLett.124.053402} {\bibfield  {journal}
  {\bibinfo  {journal} {Phys. Rev. Lett.}\ }\textbf {\bibinfo {volume} {124}},\
  \bibinfo {pages} {053402} (\bibinfo {year} {2020})}\BibitemShut {NoStop}%
\bibitem [{\citenamefont {Eid}\ \emph {et~al.}(2024)\citenamefont {Eid},
  \citenamefont {Hammond}, \citenamefont {Lavoine},\ and\ \citenamefont
  {Bourdel}}]{PRA.2024.110.043316}%
  \BibitemOpen
  \bibfield  {author} {\bibinfo {author} {\bibfnamefont {R.}~\bibnamefont
  {Eid}}, \bibinfo {author} {\bibfnamefont {A.}~\bibnamefont {Hammond}},
  \bibinfo {author} {\bibfnamefont {L.}~\bibnamefont {Lavoine}},\ and\ \bibinfo
  {author} {\bibfnamefont {T.}~\bibnamefont {Bourdel}},\ }\href
  {https://doi.org/10.1103/PhysRevA.110.043316} {\bibfield  {journal} {\bibinfo
   {journal} {Phys. Rev. A}\ }\textbf {\bibinfo {volume} {110}},\ \bibinfo
  {pages} {043316} (\bibinfo {year} {2024})}\BibitemShut {NoStop}%
\bibitem [{\citenamefont {Li}\ \emph {et~al.}(2016)\citenamefont {Li},
  \citenamefont {Wyart}, \citenamefont {Dulieu}, \citenamefont {Nascimbène},\
  and\ \citenamefont {Lepers}}]{JPB.2017.50.014005}%
  \BibitemOpen
  \bibfield  {author} {\bibinfo {author} {\bibfnamefont {H.}~\bibnamefont
  {Li}}, \bibinfo {author} {\bibfnamefont {J.-F.}\ \bibnamefont {Wyart}},
  \bibinfo {author} {\bibfnamefont {O.}~\bibnamefont {Dulieu}}, \bibinfo
  {author} {\bibfnamefont {S.}~\bibnamefont {Nascimbène}},\ and\ \bibinfo
  {author} {\bibfnamefont {M.}~\bibnamefont {Lepers}},\ }\href
  {https://doi.org/10.1088/1361-6455/50/1/014005} {\bibfield  {journal}
  {\bibinfo  {journal} {Journal of Physics B: Atomic, Molecular and Optical
  Physics}\ }\textbf {\bibinfo {volume} {50}},\ \bibinfo {pages} {014005}
  (\bibinfo {year} {2016})}\BibitemShut {NoStop}%
\bibitem [{\citenamefont {Jensen}\ and\ \citenamefont
  {Koppe}(1971)}]{jensen1971}%
  \BibitemOpen
  \bibfield  {author} {\bibinfo {author} {\bibfnamefont {H.}~\bibnamefont
  {Jensen}}\ and\ \bibinfo {author} {\bibfnamefont {H.}~\bibnamefont {Koppe}},\
  }\href {https://doi.org/10.1016/0003-4916(71)90038-8} {\bibfield  {journal}
  {\bibinfo  {journal} {Ann. Phys.}\ }\textbf {\bibinfo {volume} {63}},\
  \bibinfo {pages} {586} (\bibinfo {year} {1971})}\BibitemShut {NoStop}%
\bibitem [{\citenamefont {da~Costa}(1981)}]{dacosta1981}%
  \BibitemOpen
  \bibfield  {author} {\bibinfo {author} {\bibfnamefont {R.~C.~T.}\
  \bibnamefont {da~Costa}},\ }\href {https://doi.org/10.1103/PhysRevA.23.1982}
  {\bibfield  {journal} {\bibinfo  {journal} {Phys. Rev. A}\ }\textbf {\bibinfo
  {volume} {23}},\ \bibinfo {pages} {1982} (\bibinfo {year}
  {1981})}\BibitemShut {NoStop}%
\bibitem [{\citenamefont {Szameit}\ and\ \citenamefont
  {Nolte}(2010)}]{Szameit2010}%
  \BibitemOpen
  \bibfield  {author} {\bibinfo {author} {\bibfnamefont {A.}~\bibnamefont
  {Szameit}}\ and\ \bibinfo {author} {\bibfnamefont {S.}~\bibnamefont
  {Nolte}},\ }\href {https://doi.org/10.1088/0953-4075/43/16/163001} {\bibfield
   {journal} {\bibinfo  {journal} {J.~Phys.~B}\ }\textbf {\bibinfo {volume}
  {43}},\ \bibinfo {pages} {163001} (\bibinfo {year} {2010})}\BibitemShut
  {NoStop}%
\bibitem [{\citenamefont {Ortix}(2019)}]{CostaCord2019}%
  \BibitemOpen
  \bibfield  {author} {\bibinfo {author} {\bibfnamefont {A.}~\bibnamefont
  {Ortix}},\ }\href {https://doi.org/10.1016/j.physrep.2020.07.001} {\bibfield
  {journal} {\bibinfo  {journal} {Phys. Rep.}\ }\textbf {\bibinfo {volume}
  {886}},\ \bibinfo {pages} {1} (\bibinfo {year} {2019})}\BibitemShut {NoStop}%
\bibitem [{\citenamefont {Amitani}\ and\ \citenamefont
  {Nishida}(2023)}]{AoP.2023.448.169181}%
  \BibitemOpen
  \bibfield  {author} {\bibinfo {author} {\bibfnamefont {T.}~\bibnamefont
  {Amitani}}\ and\ \bibinfo {author} {\bibfnamefont {Y.}~\bibnamefont
  {Nishida}},\ }\href
  {https://doi.org/https://doi.org/10.1016/j.aop.2022.169181} {\bibfield
  {journal} {\bibinfo  {journal} {Annals of Physics}\ }\textbf {\bibinfo
  {volume} {448}},\ \bibinfo {pages} {169181} (\bibinfo {year}
  {2023})}\BibitemShut {NoStop}%
\bibitem [{\citenamefont {Bratman}\ \emph {et~al.}(2000)\citenamefont
  {Bratman}, \citenamefont {Cross}, \citenamefont {Denisov}, \citenamefont
  {He}, \citenamefont {Phelps}, \citenamefont {Ronald}, \citenamefont
  {Samsonov}, \citenamefont {Whyte},\ and\ \citenamefont
  {Young}}]{PRL.2000.84.2746}%
  \BibitemOpen
  \bibfield  {author} {\bibinfo {author} {\bibfnamefont {V.~L.}\ \bibnamefont
  {Bratman}}, \bibinfo {author} {\bibfnamefont {A.~W.}\ \bibnamefont {Cross}},
  \bibinfo {author} {\bibfnamefont {G.~G.}\ \bibnamefont {Denisov}}, \bibinfo
  {author} {\bibfnamefont {W.}~\bibnamefont {He}}, \bibinfo {author}
  {\bibfnamefont {A.~D.~R.}\ \bibnamefont {Phelps}}, \bibinfo {author}
  {\bibfnamefont {K.}~\bibnamefont {Ronald}}, \bibinfo {author} {\bibfnamefont
  {S.~V.}\ \bibnamefont {Samsonov}}, \bibinfo {author} {\bibfnamefont {C.~G.}\
  \bibnamefont {Whyte}},\ and\ \bibinfo {author} {\bibfnamefont {A.~R.}\
  \bibnamefont {Young}},\ }\href {https://doi.org/10.1103/PhysRevLett.84.2746}
  {\bibfield  {journal} {\bibinfo  {journal} {Phys. Rev. Lett.}\ }\textbf
  {\bibinfo {volume} {84}},\ \bibinfo {pages} {2746} (\bibinfo {year}
  {2000})}\BibitemShut {NoStop}%
\bibitem [{\citenamefont {Zhang}\ \emph {et~al.}(2022)\citenamefont {Zhang},
  \citenamefont {Easton}, \citenamefont {Donaldson}, \citenamefont {Whyte},\
  and\ \citenamefont {Cross}}]{IEEE.2022.69.3427}%
  \BibitemOpen
  \bibfield  {author} {\bibinfo {author} {\bibfnamefont {L.}~\bibnamefont
  {Zhang}}, \bibinfo {author} {\bibfnamefont {J.}~\bibnamefont {Easton}},
  \bibinfo {author} {\bibfnamefont {C.~R.}\ \bibnamefont {Donaldson}}, \bibinfo
  {author} {\bibfnamefont {C.~G.}\ \bibnamefont {Whyte}},\ and\ \bibinfo
  {author} {\bibfnamefont {A.~W.}\ \bibnamefont {Cross}},\ }\href
  {https://doi.org/10.1109/TED.2022.3170279} {\bibfield  {journal} {\bibinfo
  {journal} {IEEE Transactions on Electron Devices}\ }\textbf {\bibinfo
  {volume} {69}},\ \bibinfo {pages} {3427} (\bibinfo {year}
  {2022})}\BibitemShut {NoStop}%
\bibitem [{\citenamefont {Rozental'}\ \emph {et~al.}(2022)\citenamefont
  {Rozental'}, \citenamefont {Bogdashov}, \citenamefont {Gachev},\ and\
  \citenamefont {Samsonov}}]{RQE.2022.65.183}%
  \BibitemOpen
  \bibfield  {author} {\bibinfo {author} {\bibfnamefont {R.~M.}\ \bibnamefont
  {Rozental'}}, \bibinfo {author} {\bibfnamefont {A.~A.}\ \bibnamefont
  {Bogdashov}}, \bibinfo {author} {\bibfnamefont {I.~G.}\ \bibnamefont
  {Gachev}},\ and\ \bibinfo {author} {\bibfnamefont {S.~V.}\ \bibnamefont
  {Samsonov}},\ }\href {https://doi.org/10.1007/s11141-023-10204-8} {\bibfield
  {journal} {\bibinfo  {journal} {Radiophysics and Quantum Electronics}\
  }\textbf {\bibinfo {volume} {65}},\ \bibinfo {pages} {183} (\bibinfo {year}
  {2022})}\BibitemShut {NoStop}%
\bibitem [{\citenamefont {Rechtsman}\ \emph {et~al.}(2013)\citenamefont
  {Rechtsman}, \citenamefont {Zeuner}, \citenamefont {Plotnik}, \citenamefont
  {Lumer}, \citenamefont {Nolte}, \citenamefont {Segev},\ and\ \citenamefont
  {Szameit}}]{Rechtsman2013}%
  \BibitemOpen
  \bibfield  {author} {\bibinfo {author} {\bibfnamefont {M.~C.}\ \bibnamefont
  {Rechtsman}}, \bibinfo {author} {\bibfnamefont {J.~M.}\ \bibnamefont
  {Zeuner}}, \bibinfo {author} {\bibfnamefont {Y.}~\bibnamefont {Plotnik}},
  \bibinfo {author} {\bibfnamefont {Y.}~\bibnamefont {Lumer}}, \bibinfo
  {author} {\bibfnamefont {S.}~\bibnamefont {Nolte}}, \bibinfo {author}
  {\bibfnamefont {M.}~\bibnamefont {Segev}},\ and\ \bibinfo {author}
  {\bibfnamefont {A.}~\bibnamefont {Szameit}},\ }\href
  {https://doi.org/10.1038/nature12066} {\bibfield  {journal} {\bibinfo
  {journal} {Nature}\ }\textbf {\bibinfo {volume} {496}},\ \bibinfo {pages}
  {196} (\bibinfo {year} {2013})}\BibitemShut {NoStop}%
\bibitem [{\citenamefont {Rodríguez-Lara}\ \emph {et~al.}(2013)\citenamefont
  {Rodríguez-Lara}, \citenamefont {Aleahmad}, \citenamefont {Szameit},\ and\
  \citenamefont {Christodoulides}}]{RodriguezLara2013}%
  \BibitemOpen
  \bibfield  {author} {\bibinfo {author} {\bibfnamefont {B.~M.}\ \bibnamefont
  {Rodríguez-Lara}}, \bibinfo {author} {\bibfnamefont {P.}~\bibnamefont
  {Aleahmad}}, \bibinfo {author} {\bibfnamefont {A.}~\bibnamefont {Szameit}},\
  and\ \bibinfo {author} {\bibfnamefont {D.~N.}\ \bibnamefont
  {Christodoulides}},\ }\href {https://doi.org/10.1103/PhysRevA.87.053847}
  {\bibfield  {journal} {\bibinfo  {journal} {Phys. Rev. A}\ }\textbf {\bibinfo
  {volume} {87}},\ \bibinfo {pages} {053847} (\bibinfo {year}
  {2013})}\BibitemShut {NoStop}%
\bibitem [{\citenamefont {Henderson}\ \emph {et~al.}(2009)\citenamefont
  {Henderson}, \citenamefont {Ryu}, \citenamefont {MacCormick},\ and\
  \citenamefont {Boshier}}]{Henderson2009}%
  \BibitemOpen
  \bibfield  {author} {\bibinfo {author} {\bibfnamefont {K.}~\bibnamefont
  {Henderson}}, \bibinfo {author} {\bibfnamefont {C.}~\bibnamefont {Ryu}},
  \bibinfo {author} {\bibfnamefont {C.}~\bibnamefont {MacCormick}},\ and\
  \bibinfo {author} {\bibfnamefont {M.~G.}\ \bibnamefont {Boshier}},\ }\href
  {https://doi.org/10.1088/1367-2630/11/4/043030} {\bibfield  {journal}
  {\bibinfo  {journal} {New J. Phys.}\ }\textbf {\bibinfo {volume} {11}},\
  \bibinfo {pages} {043030} (\bibinfo {year} {2009})}\BibitemShut {NoStop}%
\bibitem [{\citenamefont {Ha}\ \emph {et~al.}(2015)\citenamefont {Ha},
  \citenamefont {LeBlanc}, \citenamefont {Luo}, \citenamefont {Corcovilos},
  \citenamefont {Cornell},\ and\ \citenamefont {Jin}}]{Ha2015}%
  \BibitemOpen
  \bibfield  {author} {\bibinfo {author} {\bibfnamefont {L.-C.}\ \bibnamefont
  {Ha}}, \bibinfo {author} {\bibfnamefont {L.}~\bibnamefont {LeBlanc}},
  \bibinfo {author} {\bibfnamefont {C.}~\bibnamefont {Luo}}, \bibinfo {author}
  {\bibfnamefont {T.}~\bibnamefont {Corcovilos}}, \bibinfo {author}
  {\bibfnamefont {E.}~\bibnamefont {Cornell}},\ and\ \bibinfo {author}
  {\bibfnamefont {D.}~\bibnamefont {Jin}},\ }\href
  {https://doi.org/10.1103/PhysRevLett.114.055301} {\bibfield  {journal}
  {\bibinfo  {journal} {Phys. Rev. Lett.}\ }\textbf {\bibinfo {volume} {114}},\
  \bibinfo {pages} {055301} (\bibinfo {year} {2015})}\BibitemShut {NoStop}%
\bibitem [{\citenamefont {Prinz}\ \emph {et~al.}(2000)\citenamefont {Prinz},
  \citenamefont {Gr\"utzmacher}, \citenamefont {Beyer}, \citenamefont {David},
  \citenamefont {Ketterer},\ and\ \citenamefont {Deckardt}}]{Prinz2000}%
  \BibitemOpen
  \bibfield  {author} {\bibinfo {author} {\bibfnamefont {V.~Y.}\ \bibnamefont
  {Prinz}}, \bibinfo {author} {\bibfnamefont {D.}~\bibnamefont
  {Gr\"utzmacher}}, \bibinfo {author} {\bibfnamefont {A.}~\bibnamefont
  {Beyer}}, \bibinfo {author} {\bibfnamefont {C.}~\bibnamefont {David}},
  \bibinfo {author} {\bibfnamefont {B.}~\bibnamefont {Ketterer}},\ and\
  \bibinfo {author} {\bibfnamefont {E.}~\bibnamefont {Deckardt}},\ }\href
  {https://doi.org/10.1016/S1386-9477(99)00417-7} {\bibfield  {journal}
  {\bibinfo  {journal} {Physica E}\ }\textbf {\bibinfo {volume} {6}},\ \bibinfo
  {pages} {828} (\bibinfo {year} {2000})}\BibitemShut {NoStop}%
\bibitem [{\citenamefont {Silva~Netto}\ and\ \citenamefont
  {Furtado}(2008)}]{JPCM.2008.20.125209}%
  \BibitemOpen
  \bibfield  {author} {\bibinfo {author} {\bibfnamefont {A.~L.}\ \bibnamefont
  {Silva~Netto}}\ and\ \bibinfo {author} {\bibfnamefont {C.}~\bibnamefont
  {Furtado}},\ }\href {https://doi.org/10.1088/0953-8984/20/12/125209}
  {\bibfield  {journal} {\bibinfo  {journal} {Journal of Physics: Condensed
  Matter}\ }\textbf {\bibinfo {volume} {20}},\ \bibinfo {pages} {125209}
  (\bibinfo {year} {2008})}\BibitemShut {NoStop}%
\bibitem [{Note1()}]{Note1}%
  \BibitemOpen
  \bibinfo {note} {Throughout this Section, spin is neglected; the probability
  current is defined in the standard gauge-invariant way~\cite
  {sakurai}.}\BibitemShut {Stop}%
\bibitem [{\citenamefont {Sakurai}\ and\ \citenamefont
  {Napolitano}(2011)}]{sakurai}%
  \BibitemOpen
  \bibfield  {author} {\bibinfo {author} {\bibfnamefont {J.~J.}\ \bibnamefont
  {Sakurai}}\ and\ \bibinfo {author} {\bibfnamefont {J.}~\bibnamefont
  {Napolitano}},\ }\href {https://doi.org/10.1063/1.1582078} {\emph {\bibinfo
  {title} {Modern Quantum Mechanics}}},\ \bibinfo {edition} {2nd}\ ed.\
  (\bibinfo  {publisher} {Pearson},\ \bibinfo {year} {2011})\BibitemShut
  {NoStop}%
\bibitem [{Note2()}]{Note2}%
  \BibitemOpen
  \bibinfo {note} {Standard optics derivations can be found in Saleh and Teich,
  Fundamentals of Photonics, 3rd ed. (Wiley, 2019).}\BibitemShut {Stop}%
\bibitem [{\citenamefont {Li}\ \emph {et~al.}(2024)\citenamefont {Li},
  \citenamefont {Chen}, \citenamefont {Chen}, \citenamefont {Luo},\ and\
  \citenamefont {Zhao}}]{PRApp.2024.22.014047}%
  \BibitemOpen
  \bibfield  {author} {\bibinfo {author} {\bibfnamefont {P.}~\bibnamefont
  {Li}}, \bibinfo {author} {\bibfnamefont {W.}~\bibnamefont {Chen}}, \bibinfo
  {author} {\bibfnamefont {J.}~\bibnamefont {Chen}}, \bibinfo {author}
  {\bibfnamefont {W.}~\bibnamefont {Luo}},\ and\ \bibinfo {author}
  {\bibfnamefont {D.}~\bibnamefont {Zhao}},\ }\href
  {https://doi.org/10.1103/PhysRevApplied.22.014047} {\bibfield  {journal}
  {\bibinfo  {journal} {Phys. Rev. Appl.}\ }\textbf {\bibinfo {volume} {22}},\
  \bibinfo {pages} {014047} (\bibinfo {year} {2024})}\BibitemShut {NoStop}%
\bibitem [{\citenamefont {Liang}\ \emph {et~al.}(2024)\citenamefont {Liang},
  \citenamefont {Dong}, \citenamefont {Pan}, \citenamefont {Wang},
  \citenamefont {Li}, \citenamefont {Yang}, \citenamefont {Yi},\ and\
  \citenamefont {Yan}}]{NP.2024.20.1738}%
  \BibitemOpen
  \bibfield  {author} {\bibinfo {author} {\bibfnamefont {Q.}~\bibnamefont
  {Liang}}, \bibinfo {author} {\bibfnamefont {Z.}~\bibnamefont {Dong}},
  \bibinfo {author} {\bibfnamefont {J.-S.}\ \bibnamefont {Pan}}, \bibinfo
  {author} {\bibfnamefont {H.}~\bibnamefont {Wang}}, \bibinfo {author}
  {\bibfnamefont {H.}~\bibnamefont {Li}}, \bibinfo {author} {\bibfnamefont
  {Z.}~\bibnamefont {Yang}}, \bibinfo {author} {\bibfnamefont {W.}~\bibnamefont
  {Yi}},\ and\ \bibinfo {author} {\bibfnamefont {B.}~\bibnamefont {Yan}},\
  }\href {https://doi.org/10.1038/s41567-024-02644-4} {\bibfield  {journal}
  {\bibinfo  {journal} {Nature Physics}\ }\textbf {\bibinfo {volume} {20}},\
  \bibinfo {pages} {1738} (\bibinfo {year} {2024})}\BibitemShut {NoStop}%
\bibitem [{\citenamefont {Lu}\ \emph {et~al.}(2025)\citenamefont {Lu},
  \citenamefont {Wang}, \citenamefont {Zou},\ and\ \citenamefont
  {Xiang}}]{PRApp.2025.23.054080}%
  \BibitemOpen
  \bibfield  {author} {\bibinfo {author} {\bibfnamefont {X.-L.}\ \bibnamefont
  {Lu}}, \bibinfo {author} {\bibfnamefont {F.-H.}\ \bibnamefont {Wang}},
  \bibinfo {author} {\bibfnamefont {J.-J.}\ \bibnamefont {Zou}},\ and\ \bibinfo
  {author} {\bibfnamefont {Z.-L.}\ \bibnamefont {Xiang}},\ }\href
  {https://doi.org/10.1103/PhysRevApplied.23.054080} {\bibfield  {journal}
  {\bibinfo  {journal} {Phys. Rev. Appl.}\ }\textbf {\bibinfo {volume} {23}},\
  \bibinfo {pages} {054080} (\bibinfo {year} {2025})}\BibitemShut {NoStop}%
\bibitem [{\citenamefont {Yannopapas}(2025)}]{photonics.2025.12.612}%
  \BibitemOpen
  \bibfield  {author} {\bibinfo {author} {\bibfnamefont {V.}~\bibnamefont
  {Yannopapas}},\ }\bibfield  {journal} {\bibinfo  {journal} {Photonics}\
  }\textbf {\bibinfo {volume} {12}},\ \href
  {https://doi.org/10.3390/photonics12060612} {10.3390/photonics12060612}
  (\bibinfo {year} {2025})\BibitemShut {NoStop}%
\bibitem [{\citenamefont {Zhang}\ \emph {et~al.}(2016)\citenamefont {Zhang},
  \citenamefont {Hsu},\ and\ \citenamefont {Liu}}]{PRB.2016.93.235315}%
  \BibitemOpen
  \bibfield  {author} {\bibinfo {author} {\bibfnamefont {R.-X.}\ \bibnamefont
  {Zhang}}, \bibinfo {author} {\bibfnamefont {H.-C.}\ \bibnamefont {Hsu}},\
  and\ \bibinfo {author} {\bibfnamefont {C.-X.}\ \bibnamefont {Liu}},\ }\href
  {https://doi.org/10.1103/PhysRevB.93.235315} {\bibfield  {journal} {\bibinfo
  {journal} {Phys. Rev. B}\ }\textbf {\bibinfo {volume} {93}},\ \bibinfo
  {pages} {235315} (\bibinfo {year} {2016})}\BibitemShut {NoStop}%
\bibitem [{\citenamefont {Goldman}\ \emph {et~al.}(2016)\citenamefont
  {Goldman}, \citenamefont {Jotzu}, \citenamefont {Messer}, \citenamefont
  {G\"org}, \citenamefont {Desbuquois},\ and\ \citenamefont
  {Esslinger}}]{PRA.2016.94.043611}%
  \BibitemOpen
  \bibfield  {author} {\bibinfo {author} {\bibfnamefont {N.}~\bibnamefont
  {Goldman}}, \bibinfo {author} {\bibfnamefont {G.}~\bibnamefont {Jotzu}},
  \bibinfo {author} {\bibfnamefont {M.}~\bibnamefont {Messer}}, \bibinfo
  {author} {\bibfnamefont {F.}~\bibnamefont {G\"org}}, \bibinfo {author}
  {\bibfnamefont {R.}~\bibnamefont {Desbuquois}},\ and\ \bibinfo {author}
  {\bibfnamefont {T.}~\bibnamefont {Esslinger}},\ }\href
  {https://doi.org/10.1103/PhysRevA.94.043611} {\bibfield  {journal} {\bibinfo
  {journal} {Phys. Rev. A}\ }\textbf {\bibinfo {volume} {94}},\ \bibinfo
  {pages} {043611} (\bibinfo {year} {2016})}\BibitemShut {NoStop}%
\bibitem [{\citenamefont {Beenakker}(2024)}]{PRB.2024.110.165421}%
  \BibitemOpen
  \bibfield  {author} {\bibinfo {author} {\bibfnamefont {C.~W.~J.}\
  \bibnamefont {Beenakker}},\ }\href
  {https://doi.org/10.1103/PhysRevB.110.165421} {\bibfield  {journal} {\bibinfo
   {journal} {Phys. Rev. B}\ }\textbf {\bibinfo {volume} {110}},\ \bibinfo
  {pages} {165421} (\bibinfo {year} {2024})}\BibitemShut {NoStop}%
\bibitem [{\citenamefont {Ziegler}\ and\ \citenamefont
  {Kezerashvili}(2025)}]{PRB.2025.111.L140507}%
  \BibitemOpen
  \bibfield  {author} {\bibinfo {author} {\bibfnamefont {K.}~\bibnamefont
  {Ziegler}}\ and\ \bibinfo {author} {\bibfnamefont {R.~Y.}\ \bibnamefont
  {Kezerashvili}},\ }\href {https://doi.org/10.1103/PhysRevB.111.L140507}
  {\bibfield  {journal} {\bibinfo  {journal} {Phys. Rev. B}\ }\textbf {\bibinfo
  {volume} {111}},\ \bibinfo {pages} {L140507} (\bibinfo {year}
  {2025})}\BibitemShut {NoStop}%
\bibitem [{\citenamefont {Hasan}\ and\ \citenamefont
  {Kane}(2010)}]{RMP.2010.82.3045}%
  \BibitemOpen
  \bibfield  {author} {\bibinfo {author} {\bibfnamefont {M.~Z.}\ \bibnamefont
  {Hasan}}\ and\ \bibinfo {author} {\bibfnamefont {C.~L.}\ \bibnamefont
  {Kane}},\ }\href {https://doi.org/10.1103/RevModPhys.82.3045} {\bibfield
  {journal} {\bibinfo  {journal} {Reviews of Modern Physics}\ }\textbf
  {\bibinfo {volume} {82}},\ \bibinfo {pages} {3045} (\bibinfo {year}
  {2010})}\BibitemShut {NoStop}%
\bibitem [{\citenamefont {Ozawa}\ \emph {et~al.}(2019)\citenamefont {Ozawa},
  \citenamefont {Price}, \citenamefont {Amo}, \citenamefont {Goldman},
  \citenamefont {Hafezi}, \citenamefont {Lu}, \citenamefont {Rechtsman},
  \citenamefont {Schuster}, \citenamefont {Simon}, \citenamefont {Zilberberg}
  \emph {et~al.}}]{RMP.2019.91.015006}%
  \BibitemOpen
  \bibfield  {author} {\bibinfo {author} {\bibfnamefont {T.}~\bibnamefont
  {Ozawa}}, \bibinfo {author} {\bibfnamefont {H.~M.}\ \bibnamefont {Price}},
  \bibinfo {author} {\bibfnamefont {A.}~\bibnamefont {Amo}}, \bibinfo {author}
  {\bibfnamefont {N.}~\bibnamefont {Goldman}}, \bibinfo {author} {\bibfnamefont
  {M.}~\bibnamefont {Hafezi}}, \bibinfo {author} {\bibfnamefont
  {L.}~\bibnamefont {Lu}}, \bibinfo {author} {\bibfnamefont {M.~C.}\
  \bibnamefont {Rechtsman}}, \bibinfo {author} {\bibfnamefont {D.}~\bibnamefont
  {Schuster}}, \bibinfo {author} {\bibfnamefont {J.}~\bibnamefont {Simon}},
  \bibinfo {author} {\bibfnamefont {O.}~\bibnamefont {Zilberberg}}, \emph
  {et~al.},\ }\href {https://doi.org/10.1103/RevModPhys.91.015006} {\bibfield
  {journal} {\bibinfo  {journal} {Reviews of Modern Physics}\ }\textbf
  {\bibinfo {volume} {91}},\ \bibinfo {pages} {015006} (\bibinfo {year}
  {2019})}\BibitemShut {NoStop}%
\bibitem [{\citenamefont {Wan}\ \emph {et~al.}(2011)\citenamefont {Wan},
  \citenamefont {Turner}, \citenamefont {Vishwanath},\ and\ \citenamefont
  {Savrasov}}]{PRB.2011.83.205101}%
  \BibitemOpen
  \bibfield  {author} {\bibinfo {author} {\bibfnamefont {X.}~\bibnamefont
  {Wan}}, \bibinfo {author} {\bibfnamefont {A.~M.}\ \bibnamefont {Turner}},
  \bibinfo {author} {\bibfnamefont {A.}~\bibnamefont {Vishwanath}},\ and\
  \bibinfo {author} {\bibfnamefont {S.~Y.}\ \bibnamefont {Savrasov}},\ }\href
  {https://doi.org/10.1103/PhysRevB.83.205101} {\bibfield  {journal} {\bibinfo
  {journal} {Phys. Rev. B}\ }\textbf {\bibinfo {volume} {83}},\ \bibinfo
  {pages} {205101} (\bibinfo {year} {2011})}\BibitemShut {NoStop}%
\bibitem [{\citenamefont {Po}\ \emph {et~al.}(2017)\citenamefont {Po},
  \citenamefont {Vishwanath},\ and\ \citenamefont
  {Watanabe}}]{Nature.2017.8.50}%
  \BibitemOpen
  \bibfield  {author} {\bibinfo {author} {\bibfnamefont {H.~C.}\ \bibnamefont
  {Po}}, \bibinfo {author} {\bibfnamefont {A.}~\bibnamefont {Vishwanath}},\
  and\ \bibinfo {author} {\bibfnamefont {H.}~\bibnamefont {Watanabe}},\ }\href
  {https://doi.org/10.1038/s41467-017-00133-2} {\bibfield  {journal} {\bibinfo
  {journal} {Nature Communications}\ }\textbf {\bibinfo {volume} {8}},\
  \bibinfo {pages} {50} (\bibinfo {year} {2017})}\BibitemShut {NoStop}%
\bibitem [{\citenamefont {Kargarian}\ and\ \citenamefont
  {Fiete}(2013)}]{PRL.2013.110.156403}%
  \BibitemOpen
  \bibfield  {author} {\bibinfo {author} {\bibfnamefont {M.}~\bibnamefont
  {Kargarian}}\ and\ \bibinfo {author} {\bibfnamefont {G.~A.}\ \bibnamefont
  {Fiete}},\ }\href {https://doi.org/10.1103/PhysRevLett.110.156403} {\bibfield
   {journal} {\bibinfo  {journal} {Phys. Rev. Lett.}\ }\textbf {\bibinfo
  {volume} {110}},\ \bibinfo {pages} {156403} (\bibinfo {year}
  {2013})}\BibitemShut {NoStop}%
\end{thebibliography}
